\documentclass[longauth]{aaEC}

\usepackage{graphicx}
\usepackage{natbib}
\usepackage{scalerel}
\usepackage{adjustbox}

\usepackage[table]{xcolor}

\usepackage{txfonts}
\usepackage[pdfencoding=auto,psdextra]{hyperref}
\hypersetup{
    colorlinks=true,
    linkcolor=blue,
    filecolor=magenta,      
    urlcolor=blue,
    citecolor=blue
}
\makeatletter
\renewcommand*\aa@pageof{, page \thepage{} of \pageref*{LastPage}}
\makeatother

\usepackage[utf8]{inputenc}

\usepackage[switch, modulo]{lineno}

\usepackage{euclid}

\defcitealias{ERONearbyGals}{H24}
\defcitealias{EROData}{C24}

\begin{document}
%
%

\title{\Euclid: Star clusters in IC 342, NGC 2403, and Holmberg II\thanks{This paper is published on behalf of the Euclid Consortium.}}    


\newcommand{\orcid}[1]{} 
\author{S.~S.~Larsen\orcid{0000-0003-0069-1203}\thanks{\email{s.larsen@astro.ru.nl}}\inst{\ref{aff1}}
\and A.~M.~N.~Ferguson\inst{\ref{aff2}}
\and J.~M.~Howell\orcid{0009-0002-2242-6515}\inst{\ref{aff2}}
\and F.~Annibali\inst{\ref{aff3}}
\and J.-C.~Cuillandre\orcid{0000-0002-3263-8645}\inst{\ref{aff4}}
\and L.~K.~Hunt\orcid{0000-0001-9162-2371}\inst{\ref{aff5}}
\and A.~Lan\c{c}on\orcid{0000-0002-7214-8296}\inst{\ref{aff6}}
\and T.~Saifollahi\orcid{0000-0002-9554-7660}\inst{\ref{aff6}}
\and D.~Massari\orcid{0000-0001-8892-4301}\inst{\ref{aff3}}
\and M.~N.~Le\orcid{0009-0003-0674-9813}\inst{\ref{aff7},\ref{aff8}}
\and N.~Aghanim\orcid{0000-0002-6688-8992}\inst{\ref{aff9}}
\and B.~Altieri\orcid{0000-0003-3936-0284}\inst{\ref{aff10}}
\and A.~Amara\inst{\ref{aff11}}
\and S.~Andreon\orcid{0000-0002-2041-8784}\inst{\ref{aff12}}
\and N.~Auricchio\orcid{0000-0003-4444-8651}\inst{\ref{aff3}}
\and C.~Baccigalupi\orcid{0000-0002-8211-1630}\inst{\ref{aff13},\ref{aff14},\ref{aff15},\ref{aff16}}
\and M.~Baldi\orcid{0000-0003-4145-1943}\inst{\ref{aff17},\ref{aff3},\ref{aff18}}
\and A.~Balestra\orcid{0000-0002-6967-261X}\inst{\ref{aff19}}
\and S.~Bardelli\orcid{0000-0002-8900-0298}\inst{\ref{aff3}}
\and P.~Battaglia\orcid{0000-0002-7337-5909}\inst{\ref{aff3}}
\and A.~Biviano\orcid{0000-0002-0857-0732}\inst{\ref{aff14},\ref{aff13}}
\and E.~Branchini\orcid{0000-0002-0808-6908}\inst{\ref{aff20},\ref{aff21},\ref{aff12}}
\and M.~Brescia\orcid{0000-0001-9506-5680}\inst{\ref{aff22},\ref{aff23}}
\and J.~Brinchmann\orcid{0000-0003-4359-8797}\inst{\ref{aff24},\ref{aff25}}
\and S.~Camera\orcid{0000-0003-3399-3574}\inst{\ref{aff26},\ref{aff27},\ref{aff28}}
\and V.~Capobianco\orcid{0000-0002-3309-7692}\inst{\ref{aff28}}
\and C.~Carbone\orcid{0000-0003-0125-3563}\inst{\ref{aff29}}
\and J.~Carretero\orcid{0000-0002-3130-0204}\inst{\ref{aff30},\ref{aff31}}
\and S.~Casas\orcid{0000-0002-4751-5138}\inst{\ref{aff32}}
\and M.~Castellano\orcid{0000-0001-9875-8263}\inst{\ref{aff33}}
\and G.~Castignani\orcid{0000-0001-6831-0687}\inst{\ref{aff3}}
\and S.~Cavuoti\orcid{0000-0002-3787-4196}\inst{\ref{aff23},\ref{aff34}}
\and K.~C.~Chambers\orcid{0000-0001-6965-7789}\inst{\ref{aff35}}
\and A.~Cimatti\inst{\ref{aff36}}
\and C.~Colodro-Conde\inst{\ref{aff7}}
\and G.~Congedo\orcid{0000-0003-2508-0046}\inst{\ref{aff2}}
\and C.~J.~Conselice\orcid{0000-0003-1949-7638}\inst{\ref{aff37}}
\and L.~Conversi\orcid{0000-0002-6710-8476}\inst{\ref{aff38},\ref{aff10}}
\and Y.~Copin\orcid{0000-0002-5317-7518}\inst{\ref{aff39}}
\and F.~Courbin\orcid{0000-0003-0758-6510}\inst{\ref{aff40},\ref{aff41}}
\and H.~M.~Courtois\orcid{0000-0003-0509-1776}\inst{\ref{aff42}}
\and M.~Cropper\orcid{0000-0003-4571-9468}\inst{\ref{aff43}}
\and A.~Da~Silva\orcid{0000-0002-6385-1609}\inst{\ref{aff44},\ref{aff45}}
\and H.~Degaudenzi\orcid{0000-0002-5887-6799}\inst{\ref{aff46}}
\and G.~De~Lucia\orcid{0000-0002-6220-9104}\inst{\ref{aff14}}
\and H.~Dole\orcid{0000-0002-9767-3839}\inst{\ref{aff9}}
\and M.~Douspis\orcid{0000-0003-4203-3954}\inst{\ref{aff9}}
\and F.~Dubath\orcid{0000-0002-6533-2810}\inst{\ref{aff46}}
\and X.~Dupac\inst{\ref{aff10}}
\and S.~Dusini\orcid{0000-0002-1128-0664}\inst{\ref{aff47}}
\and S.~Escoffier\orcid{0000-0002-2847-7498}\inst{\ref{aff48}}
\and M.~Fabricius\orcid{0000-0002-7025-6058}\inst{\ref{aff49},\ref{aff50}}
\and M.~Farina\orcid{0000-0002-3089-7846}\inst{\ref{aff51}}
\and F.~Faustini\orcid{0000-0001-6274-5145}\inst{\ref{aff33},\ref{aff52}}
\and S.~Ferriol\inst{\ref{aff39}}
\and S.~Fotopoulou\orcid{0000-0002-9686-254X}\inst{\ref{aff53}}
\and M.~Frailis\orcid{0000-0002-7400-2135}\inst{\ref{aff14}}
\and E.~Franceschi\orcid{0000-0002-0585-6591}\inst{\ref{aff3}}
\and S.~Galeotta\orcid{0000-0002-3748-5115}\inst{\ref{aff14}}
\and K.~George\orcid{0000-0002-1734-8455}\inst{\ref{aff50}}
\and B.~Gillis\orcid{0000-0002-4478-1270}\inst{\ref{aff2}}
\and C.~Giocoli\orcid{0000-0002-9590-7961}\inst{\ref{aff3},\ref{aff18}}
\and P.~G\'omez-Alvarez\orcid{0000-0002-8594-5358}\inst{\ref{aff54},\ref{aff10}}
\and J.~Gracia-Carpio\inst{\ref{aff49}}
\and A.~Grazian\orcid{0000-0002-5688-0663}\inst{\ref{aff19}}
\and F.~Grupp\inst{\ref{aff49},\ref{aff50}}
\and S.~V.~H.~Haugan\orcid{0000-0001-9648-7260}\inst{\ref{aff55}}
\and H.~Hoekstra\orcid{0000-0002-0641-3231}\inst{\ref{aff56}}
\and W.~Holmes\inst{\ref{aff57}}
\and F.~Hormuth\inst{\ref{aff58}}
\and A.~Hornstrup\orcid{0000-0002-3363-0936}\inst{\ref{aff59},\ref{aff60}}
\and K.~Jahnke\orcid{0000-0003-3804-2137}\inst{\ref{aff61}}
\and M.~Jhabvala\inst{\ref{aff62}}
\and E.~Keih\"anen\orcid{0000-0003-1804-7715}\inst{\ref{aff63}}
\and S.~Kermiche\orcid{0000-0002-0302-5735}\inst{\ref{aff48}}
\and A.~Kiessling\orcid{0000-0002-2590-1273}\inst{\ref{aff57}}
\and R.~Kohley\inst{\ref{aff10}}
\and B.~Kubik\orcid{0009-0006-5823-4880}\inst{\ref{aff39}}
\and K.~Kuijken\orcid{0000-0002-3827-0175}\inst{\ref{aff56}}
\and M.~K\"ummel\orcid{0000-0003-2791-2117}\inst{\ref{aff50}}
\and M.~Kunz\orcid{0000-0002-3052-7394}\inst{\ref{aff64}}
\and H.~Kurki-Suonio\orcid{0000-0002-4618-3063}\inst{\ref{aff65},\ref{aff66}}
\and A.~M.~C.~Le~Brun\orcid{0000-0002-0936-4594}\inst{\ref{aff67}}
\and S.~Ligori\orcid{0000-0003-4172-4606}\inst{\ref{aff28}}
\and P.~B.~Lilje\orcid{0000-0003-4324-7794}\inst{\ref{aff55}}
\and V.~Lindholm\orcid{0000-0003-2317-5471}\inst{\ref{aff65},\ref{aff66}}
\and I.~Lloro\orcid{0000-0001-5966-1434}\inst{\ref{aff68}}
\and D.~Maino\inst{\ref{aff69},\ref{aff29},\ref{aff70}}
\and E.~Maiorano\orcid{0000-0003-2593-4355}\inst{\ref{aff3}}
\and O.~Mansutti\orcid{0000-0001-5758-4658}\inst{\ref{aff14}}
\and O.~Marggraf\orcid{0000-0001-7242-3852}\inst{\ref{aff71}}
\and K.~Markovic\orcid{0000-0001-6764-073X}\inst{\ref{aff57}}
\and M.~Martinelli\orcid{0000-0002-6943-7732}\inst{\ref{aff33},\ref{aff72}}
\and N.~Martinet\orcid{0000-0003-2786-7790}\inst{\ref{aff73}}
\and F.~Marulli\orcid{0000-0002-8850-0303}\inst{\ref{aff74},\ref{aff3},\ref{aff18}}
\and R.~Massey\orcid{0000-0002-6085-3780}\inst{\ref{aff75}}
\and E.~Medinaceli\orcid{0000-0002-4040-7783}\inst{\ref{aff3}}
\and S.~Mei\orcid{0000-0002-2849-559X}\inst{\ref{aff76},\ref{aff77}}
\and Y.~Mellier\inst{\ref{aff78},\ref{aff79}}
\and M.~Meneghetti\orcid{0000-0003-1225-7084}\inst{\ref{aff3},\ref{aff18}}
\and G.~Meylan\inst{\ref{aff80}}
\and A.~Mora\orcid{0000-0002-1922-8529}\inst{\ref{aff81}}
\and M.~Moresco\orcid{0000-0002-7616-7136}\inst{\ref{aff74},\ref{aff3}}
\and L.~Moscardini\orcid{0000-0002-3473-6716}\inst{\ref{aff74},\ref{aff3},\ref{aff18}}
\and C.~Neissner\orcid{0000-0001-8524-4968}\inst{\ref{aff82},\ref{aff31}}
\and S.-M.~Niemi\orcid{0009-0005-0247-0086}\inst{\ref{aff83}}
\and C.~Padilla\orcid{0000-0001-7951-0166}\inst{\ref{aff82}}
\and S.~Paltani\orcid{0000-0002-8108-9179}\inst{\ref{aff46}}
\and F.~Pasian\orcid{0000-0002-4869-3227}\inst{\ref{aff14}}
\and K.~Pedersen\inst{\ref{aff84}}
\and W.~J.~Percival\orcid{0000-0002-0644-5727}\inst{\ref{aff85},\ref{aff86},\ref{aff87}}
\and V.~Pettorino\inst{\ref{aff83}}
\and S.~Pires\orcid{0000-0002-0249-2104}\inst{\ref{aff4}}
\and G.~Polenta\orcid{0000-0003-4067-9196}\inst{\ref{aff52}}
\and M.~Poncet\inst{\ref{aff88}}
\and L.~A.~Popa\inst{\ref{aff89}}
\and L.~Pozzetti\orcid{0000-0001-7085-0412}\inst{\ref{aff3}}
\and F.~Raison\orcid{0000-0002-7819-6918}\inst{\ref{aff49}}
\and A.~Renzi\orcid{0000-0001-9856-1970}\inst{\ref{aff90},\ref{aff47}}
\and J.~Rhodes\orcid{0000-0002-4485-8549}\inst{\ref{aff57}}
\and G.~Riccio\inst{\ref{aff23}}
\and E.~Romelli\orcid{0000-0003-3069-9222}\inst{\ref{aff14}}
\and M.~Roncarelli\orcid{0000-0001-9587-7822}\inst{\ref{aff3}}
\and R.~Saglia\orcid{0000-0003-0378-7032}\inst{\ref{aff50},\ref{aff49}}
\and Z.~Sakr\orcid{0000-0002-4823-3757}\inst{\ref{aff91},\ref{aff92},\ref{aff93}}
\and D.~Sapone\orcid{0000-0001-7089-4503}\inst{\ref{aff94}}
\and B.~Sartoris\orcid{0000-0003-1337-5269}\inst{\ref{aff50},\ref{aff14}}
\and J.~A.~Schewtschenko\orcid{0000-0002-4913-6393}\inst{\ref{aff2}}
\and M.~Schirmer\orcid{0000-0003-2568-9994}\inst{\ref{aff61}}
\and P.~Schneider\orcid{0000-0001-8561-2679}\inst{\ref{aff71}}
\and A.~Secroun\orcid{0000-0003-0505-3710}\inst{\ref{aff48}}
\and G.~Seidel\orcid{0000-0003-2907-353X}\inst{\ref{aff61}}
\and M.~Seiffert\orcid{0000-0002-7536-9393}\inst{\ref{aff57}}
\and S.~Serrano\orcid{0000-0002-0211-2861}\inst{\ref{aff95},\ref{aff96},\ref{aff97}}
\and P.~Simon\inst{\ref{aff71}}
\and C.~Sirignano\orcid{0000-0002-0995-7146}\inst{\ref{aff90},\ref{aff47}}
\and G.~Sirri\orcid{0000-0003-2626-2853}\inst{\ref{aff18}}
\and J.~Skottfelt\orcid{0000-0003-1310-8283}\inst{\ref{aff98}}
\and L.~Stanco\orcid{0000-0002-9706-5104}\inst{\ref{aff47}}
\and J.~Steinwagner\orcid{0000-0001-7443-1047}\inst{\ref{aff49}}
\and P.~Tallada-Cresp\'{i}\orcid{0000-0002-1336-8328}\inst{\ref{aff30},\ref{aff31}}
\and A.~N.~Taylor\inst{\ref{aff2}}
\and I.~Tereno\orcid{0000-0002-4537-6218}\inst{\ref{aff44},\ref{aff99}}
\and R.~Toledo-Moreo\orcid{0000-0002-2997-4859}\inst{\ref{aff100}}
\and F.~Torradeflot\orcid{0000-0003-1160-1517}\inst{\ref{aff31},\ref{aff30}}
\and A.~Tsyganov\inst{\ref{aff101}}
\and I.~Tutusaus\orcid{0000-0002-3199-0399}\inst{\ref{aff92}}
\and L.~Valenziano\orcid{0000-0002-1170-0104}\inst{\ref{aff3},\ref{aff102}}
\and J.~Valiviita\orcid{0000-0001-6225-3693}\inst{\ref{aff65},\ref{aff66}}
\and T.~Vassallo\orcid{0000-0001-6512-6358}\inst{\ref{aff50},\ref{aff14}}
\and G.~Verdoes~Kleijn\orcid{0000-0001-5803-2580}\inst{\ref{aff103}}
\and A.~Veropalumbo\orcid{0000-0003-2387-1194}\inst{\ref{aff12},\ref{aff21},\ref{aff20}}
\and Y.~Wang\orcid{0000-0002-4749-2984}\inst{\ref{aff104}}
\and J.~Weller\orcid{0000-0002-8282-2010}\inst{\ref{aff50},\ref{aff49}}
\and G.~Zamorani\orcid{0000-0002-2318-301X}\inst{\ref{aff3}}
\and F.~M.~Zerbi\inst{\ref{aff12}}
\and E.~Zucca\orcid{0000-0002-5845-8132}\inst{\ref{aff3}}
\and J.~Mart\'{i}n-Fleitas\orcid{0000-0002-8594-569X}\inst{\ref{aff81}}
\and V.~Scottez\inst{\ref{aff78},\ref{aff105}}}
										   
\institute{Department of Astrophysics/IMAPP, Radboud University, PO Box 9010, 6500 GL Nijmegen, The Netherlands\label{aff1}
\and
Institute for Astronomy, University of Edinburgh, Royal Observatory, Blackford Hill, Edinburgh EH9 3HJ, UK\label{aff2}
\and
INAF-Osservatorio di Astrofisica e Scienza dello Spazio di Bologna, Via Piero Gobetti 93/3, 40129 Bologna, Italy\label{aff3}
\and
Universit\'e Paris-Saclay, Universit\'e Paris Cit\'e, CEA, CNRS, AIM, 91191, Gif-sur-Yvette, France\label{aff4}
\and
INAF-Osservatorio Astrofisico di Arcetri, Largo E. Fermi 5, 50125, Firenze, Italy\label{aff5}
\and
Universit\'e de Strasbourg, CNRS, Observatoire astronomique de Strasbourg, UMR 7550, 67000 Strasbourg, France\label{aff6}
\and
Instituto de Astrof\'{\i}sica de Canarias, V\'{\i}a L\'actea, 38205 La Laguna, Tenerife, Spain\label{aff7}
\and
Universidad de La Laguna, Departamento de Astrof\'{\i}sica, 38206 La Laguna, Tenerife, Spain\label{aff8}
\and
Universit\'e Paris-Saclay, CNRS, Institut d'astrophysique spatiale, 91405, Orsay, France\label{aff9}
\and
ESAC/ESA, Camino Bajo del Castillo, s/n., Urb. Villafranca del Castillo, 28692 Villanueva de la Ca\~nada, Madrid, Spain\label{aff10}
\and
School of Mathematics and Physics, University of Surrey, Guildford, Surrey, GU2 7XH, UK\label{aff11}
\and
INAF-Osservatorio Astronomico di Brera, Via Brera 28, 20122 Milano, Italy\label{aff12}
\and
IFPU, Institute for Fundamental Physics of the Universe, via Beirut 2, 34151 Trieste, Italy\label{aff13}
\and
INAF-Osservatorio Astronomico di Trieste, Via G. B. Tiepolo 11, 34143 Trieste, Italy\label{aff14}
\and
INFN, Sezione di Trieste, Via Valerio 2, 34127 Trieste TS, Italy\label{aff15}
\and
SISSA, International School for Advanced Studies, Via Bonomea 265, 34136 Trieste TS, Italy\label{aff16}
\and
Dipartimento di Fisica e Astronomia, Universit\`a di Bologna, Via Gobetti 93/2, 40129 Bologna, Italy\label{aff17}
\and
INFN-Sezione di Bologna, Viale Berti Pichat 6/2, 40127 Bologna, Italy\label{aff18}
\and
INAF-Osservatorio Astronomico di Padova, Via dell'Osservatorio 5, 35122 Padova, Italy\label{aff19}
\and
Dipartimento di Fisica, Universit\`a di Genova, Via Dodecaneso 33, 16146, Genova, Italy\label{aff20}
\and
INFN-Sezione di Genova, Via Dodecaneso 33, 16146, Genova, Italy\label{aff21}
\and
Department of Physics "E. Pancini", University Federico II, Via Cinthia 6, 80126, Napoli, Italy\label{aff22}
\and
INAF-Osservatorio Astronomico di Capodimonte, Via Moiariello 16, 80131 Napoli, Italy\label{aff23}
\and
Instituto de Astrof\'isica e Ci\^encias do Espa\c{c}o, Universidade do Porto, CAUP, Rua das Estrelas, PT4150-762 Porto, Portugal\label{aff24}
\and
Faculdade de Ci\^encias da Universidade do Porto, Rua do Campo de Alegre, 4150-007 Porto, Portugal\label{aff25}
\and
Dipartimento di Fisica, Universit\`a degli Studi di Torino, Via P. Giuria 1, 10125 Torino, Italy\label{aff26}
\and
INFN-Sezione di Torino, Via P. Giuria 1, 10125 Torino, Italy\label{aff27}
\and
INAF-Osservatorio Astrofisico di Torino, Via Osservatorio 20, 10025 Pino Torinese (TO), Italy\label{aff28}
\and
INAF-IASF Milano, Via Alfonso Corti 12, 20133 Milano, Italy\label{aff29}
\and
Centro de Investigaciones Energ\'eticas, Medioambientales y Tecnol\'ogicas (CIEMAT), Avenida Complutense 40, 28040 Madrid, Spain\label{aff30}
\and
Port d'Informaci\'{o} Cient\'{i}fica, Campus UAB, C. Albareda s/n, 08193 Bellaterra (Barcelona), Spain\label{aff31}
\and
Institute for Theoretical Particle Physics and Cosmology (TTK), RWTH Aachen University, 52056 Aachen, Germany\label{aff32}
\and
INAF-Osservatorio Astronomico di Roma, Via Frascati 33, 00078 Monteporzio Catone, Italy\label{aff33}
\and
INFN section of Naples, Via Cinthia 6, 80126, Napoli, Italy\label{aff34}
\and
Institute for Astronomy, University of Hawaii, 2680 Woodlawn Drive, Honolulu, HI 96822, USA\label{aff35}
\and
Dipartimento di Fisica e Astronomia "Augusto Righi" - Alma Mater Studiorum Universit\`a di Bologna, Viale Berti Pichat 6/2, 40127 Bologna, Italy\label{aff36}
\and
Jodrell Bank Centre for Astrophysics, Department of Physics and Astronomy, University of Manchester, Oxford Road, Manchester M13 9PL, UK\label{aff37}
\and
European Space Agency/ESRIN, Largo Galileo Galilei 1, 00044 Frascati, Roma, Italy\label{aff38}
\and
Universit\'e Claude Bernard Lyon 1, CNRS/IN2P3, IP2I Lyon, UMR 5822, Villeurbanne, F-69100, France\label{aff39}
\and
Institut de Ci\`{e}ncies del Cosmos (ICCUB), Universitat de Barcelona (IEEC-UB), Mart\'{i} i Franqu\`{e}s 1, 08028 Barcelona, Spain\label{aff40}
\and
Instituci\'o Catalana de Recerca i Estudis Avan\c{c}ats (ICREA), Passeig de Llu\'{\i}s Companys 23, 08010 Barcelona, Spain\label{aff41}
\and
UCB Lyon 1, CNRS/IN2P3, IUF, IP2I Lyon, 4 rue Enrico Fermi, 69622 Villeurbanne, France\label{aff42}
\and
Mullard Space Science Laboratory, University College London, Holmbury St Mary, Dorking, Surrey RH5 6NT, UK\label{aff43}
\and
Departamento de F\'isica, Faculdade de Ci\^encias, Universidade de Lisboa, Edif\'icio C8, Campo Grande, PT1749-016 Lisboa, Portugal\label{aff44}
\and
Instituto de Astrof\'isica e Ci\^encias do Espa\c{c}o, Faculdade de Ci\^encias, Universidade de Lisboa, Campo Grande, 1749-016 Lisboa, Portugal\label{aff45}
\and
Department of Astronomy, University of Geneva, ch. d'Ecogia 16, 1290 Versoix, Switzerland\label{aff46}
\and
INFN-Padova, Via Marzolo 8, 35131 Padova, Italy\label{aff47}
\and
Aix-Marseille Universit\'e, CNRS/IN2P3, CPPM, Marseille, France\label{aff48}
\and
Max Planck Institute for Extraterrestrial Physics, Giessenbachstr. 1, 85748 Garching, Germany\label{aff49}
\and
Universit\"ats-Sternwarte M\"unchen, Fakult\"at f\"ur Physik, Ludwig-Maximilians-Universit\"at M\"unchen, Scheinerstrasse 1, 81679 M\"unchen, Germany\label{aff50}
\and
INAF-Istituto di Astrofisica e Planetologia Spaziali, via del Fosso del Cavaliere, 100, 00100 Roma, Italy\label{aff51}
\and
Space Science Data Center, Italian Space Agency, via del Politecnico snc, 00133 Roma, Italy\label{aff52}
\and
School of Physics, HH Wills Physics Laboratory, University of Bristol, Tyndall Avenue, Bristol, BS8 1TL, UK\label{aff53}
\and
FRACTAL S.L.N.E., calle Tulip\'an 2, Portal 13 1A, 28231, Las Rozas de Madrid, Spain\label{aff54}
\and
Institute of Theoretical Astrophysics, University of Oslo, P.O. Box 1029 Blindern, 0315 Oslo, Norway\label{aff55}
\and
Leiden Observatory, Leiden University, Einsteinweg 55, 2333 CC Leiden, The Netherlands\label{aff56}
\and
Jet Propulsion Laboratory, California Institute of Technology, 4800 Oak Grove Drive, Pasadena, CA, 91109, USA\label{aff57}
\and
Felix Hormuth Engineering, Goethestr. 17, 69181 Leimen, Germany\label{aff58}
\and
Technical University of Denmark, Elektrovej 327, 2800 Kgs. Lyngby, Denmark\label{aff59}
\and
Cosmic Dawn Center (DAWN), Denmark\label{aff60}
\and
Max-Planck-Institut f\"ur Astronomie, K\"onigstuhl 17, 69117 Heidelberg, Germany\label{aff61}
\and
NASA Goddard Space Flight Center, Greenbelt, MD 20771, USA\label{aff62}
\and
Department of Physics and Helsinki Institute of Physics, Gustaf H\"allstr\"omin katu 2, 00014 University of Helsinki, Finland\label{aff63}
\and
Universit\'e de Gen\`eve, D\'epartement de Physique Th\'eorique and Centre for Astroparticle Physics, 24 quai Ernest-Ansermet, CH-1211 Gen\`eve 4, Switzerland\label{aff64}
\and
Department of Physics, P.O. Box 64, 00014 University of Helsinki, Finland\label{aff65}
\and
Helsinki Institute of Physics, Gustaf H{\"a}llstr{\"o}min katu 2, University of Helsinki, Helsinki, Finland\label{aff66}
\and
Laboratoire d'etude de l'Univers et des phenomenes eXtremes, Observatoire de Paris, Universit\'e PSL, Sorbonne Universit\'e, CNRS, 92190 Meudon, France\label{aff67}
\and
SKA Observatory, Jodrell Bank, Lower Withington, Macclesfield, Cheshire SK11 9FT, UK\label{aff68}
\and
Dipartimento di Fisica "Aldo Pontremoli", Universit\`a degli Studi di Milano, Via Celoria 16, 20133 Milano, Italy\label{aff69}
\and
INFN-Sezione di Milano, Via Celoria 16, 20133 Milano, Italy\label{aff70}
\and
Universit\"at Bonn, Argelander-Institut f\"ur Astronomie, Auf dem H\"ugel 71, 53121 Bonn, Germany\label{aff71}
\and
INFN-Sezione di Roma, Piazzale Aldo Moro, 2 - c/o Dipartimento di Fisica, Edificio G. Marconi, 00185 Roma, Italy\label{aff72}
\and
Aix-Marseille Universit\'e, CNRS, CNES, LAM, Marseille, France\label{aff73}
\and
Dipartimento di Fisica e Astronomia "Augusto Righi" - Alma Mater Studiorum Universit\`a di Bologna, via Piero Gobetti 93/2, 40129 Bologna, Italy\label{aff74}
\and
Department of Physics, Institute for Computational Cosmology, Durham University, South Road, Durham, DH1 3LE, UK\label{aff75}
\and
Universit\'e Paris Cit\'e, CNRS, Astroparticule et Cosmologie, 75013 Paris, France\label{aff76}
\and
CNRS-UCB International Research Laboratory, Centre Pierre Binetruy, IRL2007, CPB-IN2P3, Berkeley, USA\label{aff77}
\and
Institut d'Astrophysique de Paris, 98bis Boulevard Arago, 75014, Paris, France\label{aff78}
\and
Institut d'Astrophysique de Paris, UMR 7095, CNRS, and Sorbonne Universit\'e, 98 bis boulevard Arago, 75014 Paris, France\label{aff79}
\and
Institute of Physics, Laboratory of Astrophysics, Ecole Polytechnique F\'ed\'erale de Lausanne (EPFL), Observatoire de Sauverny, 1290 Versoix, Switzerland\label{aff80}
\and
Aurora Technology for European Space Agency (ESA), Camino bajo del Castillo, s/n, Urbanizacion Villafranca del Castillo, Villanueva de la Ca\~nada, 28692 Madrid, Spain\label{aff81}
\and
Institut de F\'{i}sica d'Altes Energies (IFAE), The Barcelona Institute of Science and Technology, Campus UAB, 08193 Bellaterra (Barcelona), Spain\label{aff82}
\and
European Space Agency/ESTEC, Keplerlaan 1, 2201 AZ Noordwijk, The Netherlands\label{aff83}
\and
DARK, Niels Bohr Institute, University of Copenhagen, Jagtvej 155, 2200 Copenhagen, Denmark\label{aff84}
\and
Waterloo Centre for Astrophysics, University of Waterloo, Waterloo, Ontario N2L 3G1, Canada\label{aff85}
\and
Department of Physics and Astronomy, University of Waterloo, Waterloo, Ontario N2L 3G1, Canada\label{aff86}
\and
Perimeter Institute for Theoretical Physics, Waterloo, Ontario N2L 2Y5, Canada\label{aff87}
\and
Centre National d'Etudes Spatiales -- Centre spatial de Toulouse, 18 avenue Edouard Belin, 31401 Toulouse Cedex 9, France\label{aff88}
\and
Institute of Space Science, Str. Atomistilor, nr. 409 M\u{a}gurele, Ilfov, 077125, Romania\label{aff89}
\and
Dipartimento di Fisica e Astronomia "G. Galilei", Universit\`a di Padova, Via Marzolo 8, 35131 Padova, Italy\label{aff90}
\and
Institut f\"ur Theoretische Physik, University of Heidelberg, Philosophenweg 16, 69120 Heidelberg, Germany\label{aff91}
\and
Institut de Recherche en Astrophysique et Plan\'etologie (IRAP), Universit\'e de Toulouse, CNRS, UPS, CNES, 14 Av. Edouard Belin, 31400 Toulouse, France\label{aff92}
\and
Universit\'e St Joseph; Faculty of Sciences, Beirut, Lebanon\label{aff93}
\and
Departamento de F\'isica, FCFM, Universidad de Chile, Blanco Encalada 2008, Santiago, Chile\label{aff94}
\and
Institut d'Estudis Espacials de Catalunya (IEEC),  Edifici RDIT, Campus UPC, 08860 Castelldefels, Barcelona, Spain\label{aff95}
\and
Satlantis, University Science Park, Sede Bld 48940, Leioa-Bilbao, Spain\label{aff96}
\and
Institute of Space Sciences (ICE, CSIC), Campus UAB, Carrer de Can Magrans, s/n, 08193 Barcelona, Spain\label{aff97}
\and
Centre for Electronic Imaging, Open University, Walton Hall, Milton Keynes, MK7~6AA, UK\label{aff98}
\and
Instituto de Astrof\'isica e Ci\^encias do Espa\c{c}o, Faculdade de Ci\^encias, Universidade de Lisboa, Tapada da Ajuda, 1349-018 Lisboa, Portugal\label{aff99}
\and
Universidad Polit\'ecnica de Cartagena, Departamento de Electr\'onica y Tecnolog\'ia de Computadoras,  Plaza del Hospital 1, 30202 Cartagena, Spain\label{aff100}
\and
Centre for Information Technology, University of Groningen, P.O. Box 11044, 9700 CA Groningen, The Netherlands\label{aff101}
\and
INFN-Bologna, Via Irnerio 46, 40126 Bologna, Italy\label{aff102}
\and
Kapteyn Astronomical Institute, University of Groningen, PO Box 800, 9700 AV Groningen, The Netherlands\label{aff103}
\and
Infrared Processing and Analysis Center, California Institute of Technology, Pasadena, CA 91125, USA\label{aff104}
\and
ICL, Junia, Universit\'e Catholique de Lille, LITL, 59000 Lille, France\label{aff105}}    

%
%
\abstract{
We examine the star cluster populations in the three nearby (3.20--3.45\,Mpc) galaxies IC\,342,  NGC\,2403, and Holmberg\,II, observed as part of the \Euclid\ Early Release Observations programme. 
Our main focus in this paper is on old globular clusters (GCs), for which the wide field-of-view and excellent image quality of \Euclid\ offer substantial advantages over previous work. For IC\,342, in particular, this is the first study of stellar clusters other than its nuclear cluster.  
After selection based on size and magnitude criteria, followed by visual inspection, we identify 111 old ($\gtrsim1$\,Gyr) GC candidates in IC\,342, 50 in NGC\,2403 (of which 15 were previously known), and 7 in Holmberg\,II. 
In addition, a number of younger and/or intermediate-age candidates are identified. The colour distributions of GC candidates in the two larger galaxies show hints of bimodality with peaks at $(\IE-\HE)_0 = 0.36$ and 0.79 (IC\,342) and $(\IE-\HE)_0 = 0.36$ and 0.80 (NGC\,2403), corresponding to metallicities of $\mathrm{[Fe/H]}\approx-1.5$ and $\mathrm{[Fe/H]}\approx-0.5$, similar to those found for the metal-poor and metal-rich GC subpopulations in the Milky Way. 
The luminosity functions of our GC candidates exhibit an excess of relatively faint objects, relative to a canonical, approximately Gaussian GC luminosity function (GCLF). Although some contamination from background galaxies may be present in our samples, we argue that the excess may be partly real, particularly in IC\,342 where the excess objects may be similar to those previously identified in galaxies such as M101 and NGC\,6946. The specific frequency of classical old GCs in IC\,342, as determined based on the brighter half of the GCLF, appears to be unusually low with $S_N\approx0.2$--0.3.
The combined luminosity function of young and intermediate-age clusters in all three galaxies is consistent with a power-law distribution, $\mathrm{d}N/\mathrm{d}L \propto L^{-2.3\pm0.1}$. The total numbers of young clusters brighter than $M(\IE)=-8$ in NGC\,2403 and Holmberg\,II are comparable with those found in their Local Group counterparts, that is, M33 and the Small Magellanic Cloud, respectively. 
}
%
%
%
%
   \titlerunning{Star clusters in IC\,342, NGC\,2403, and Holmberg\,II }
   \authorrunning{Larsen et al.}
   
   \maketitle
%
%
%
%
   
\section{\label{sc:Intro}Introduction}

As part of the \Euclid\ Early Release Observations (\citealt{EROcite,EROData}, hereafter \citetalias{EROData}), the three galaxies IC\,342, NGC\,2403, and Holmberg\,II were observed in the context of the Nearby Galaxies Showcase programme 
(\citealt{ERONearbyGals}; hereafter \citetalias{ERONearbyGals}).
In terms of their total luminosities and overall morphologies, these galaxies are comparable to the Milky Way or M31, to M33, and to the Small Magellanic Cloud, respectively. They are located at distances of 3.20--3.45\,Mpc, close enough for the brightest red giants to be resolved in the \Euclid\ images, and for stellar clusters to appear noticeably more extended than point sources.  
In the present paper we follow up on the initial exploratory work on globular clusters (GCs) presented in \citetalias{ERONearbyGals} with a more detailed analysis of the star cluster populations in the three galaxies. An analysis of the star cluster populations in the Local Group galaxies IC\,10 and NGC\,6822, which were also included in the Showcase programme, is presented separately (Howell et al., in prep.).

A comparison of the star cluster populations of Local Group and other nearby galaxies reveals a considerable degree of diversity. In particular, the distinction between ancient (``globular'') clusters, typically associated with the spheroidal components of their parent galaxies, and younger (``open'') star clusters in star-forming discs is not always as clear as might appear to be the case in the Milky Way. 
Over the past decade, evidence has emerged that differences in the mass function of young star clusters correlate with the star-formation rate and gas surface density of the parent galaxies \citep{Larsen2009,Johnson2017,Wainer2022}, such that more massive clusters form preferentially in environments with high star-formation rates. This suggests that the ancient GCs, which must generally have been born with masses greater than $10^5\, M_\odot$ in order to survive to the present epoch, trace episodes of intense star formation in the past histories of their parent galaxies, whether these took place ``in situ'' or in smaller galaxies that were subsequently accreted.

Theoretical efforts to model the characteristics of GC populations, based upon the premise that their initial properties can be determined by applying knowledge obtained from studies of young cluster populations in various environments, have met with considerable success \citep{Choksi2018,Pfeffer2023,Reina-Campos2017,Reina-Campos2022}.  A key prediction from such modelling is that the characteristics of GC populations, such as their age-, metallicity-, and phase-space distributions, will depend on the individual hierarchical assembly histories of galaxies \citep{Pfeffer2020,Trujillo-Gomez2023}. 

There is ample empirical evidence in support of the hierarchical assembly paradigm in both the Milky Way and M31 GC systems.  
\citet{Mackey2004} estimated that about 40 of the Milky Way halo GCs have been accreted from approximately seven dwarf galaxies. Subsequent work has substantiated and refined these results, suggesting that about half of the Galactic GCs may have an accretion origin, and has even associated GCs with specific dwarf galaxies/streams, providing significant insight into the assembly history of the Milky Way \citep{Forbes2010,Massari2019,Forbes2020,Kruijssen2020,Callingham2022,Limberg2022,Belokurov2024}.
The high fraction of accreted GCs inferred from observations is in agreement with expectations from theoretical work \citep{Qu2017,Davison2020,Keller2020}.
Likewise, about half of the GCs in the outer halo of M31 appear to be associated with the halo sub-structure that is  clearly evident from observations of the field stars \citep{Mackey2010,Mackey2019,Mackey2019a}. 
The M31 GC system is significantly more populous than that of the Milky Way: the \citet[][2010 edition]{Harris1996} catalogue lists 157 GCs in the Milky Way, while M31 probably has at least three times as many \citep{Harris2013,Huxor2014,Larsen2016}. 
A similar analysis is more challenging for M33, owing to its sparser GC system. \citet{Harris2013} list an estimated total of $50\pm20$ GCs for M33, but the number of objects that have been robustly confirmed individually as ancient GCs, based either on resolved imaging \citep{Sarajedini1998,Chandar1999,Sarajedini2000,Huxor2009,Cockcroft2011} or spectroscopy \citep{Beasley2015,Larsen2022}, is much smaller. 

Beyond the Local Group, a vast literature on extragalactic GC systems has accumulated over the past several decades. 
The GC systems of spiral galaxies are particularly difficult to identify and characterise, as they are generally poorer than those of early-type galaxies and harder to pick out among a variety of potential contaminants in the discs of spirals. As in some Local Group galaxies, the distinction between GCs and young massive clusters is frequently blurred, and young clusters with masses in the range $10^5$--$10^6 \, M_\odot$ have been found in a number of actively star-forming galaxies, such as M51, M83, and NGC\,6946 \citep{Larsen1999,Larsen2001,Haas2008,Chandar2016}, in starburst dwarf galaxies, and in interacting and merging galaxies  \citep{PortegiesZwart2010,Whitmore2010,Adamo2020,Adamo2020a}. 
Generally speaking, younger clusters are expected to be bluer than their older counterparts, the GCs, but several factors conspire to make the distinction far from straight-forward in practice.  Among these is the age-metallicity degeneracy, whereby old, metal-poor stellar populations have colours similar to those of younger, more metal-rich ones \citep{Worthey1994}. Surveys such as PHANGS have demonstrated the power of high-quality multi-passband imaging for identification and rough age-dating of clusters, especially when UV imaging is included \citep{Maschmann2024}. However, even with multi-colour photometry, broad-band colours remain much more sensitive to metallicity than to age for ages greater than about a Gyr, making accurate age-dating challenging  \citep{Chies-Santos2011,Powalka2017}. 

The richness of a GC system is usually quantified by the GC specific frequency, $S_N = N_\mathrm{GC} \, 10^{0.4(M_V + 15)}$ \citep{Harris1981}, for a galaxy with $N_\mathrm{GC}$ GCs and an integrated absolute visual magnitude $M_V$. The Milky Way then has $S_N = 0.48$, while M31 has $S_N = 0.86$, which are fairly typical values for spiral galaxies \citep[assuming $M_V=-21.3$ for the Milky Way and $-21.8$ for M31, as listed by][]{Harris2013}. 
The Sombrero galaxy, M104/NGC\,4594 (type S0/Sa), has a rich GC system that can fairly easily be identified even in ground-based data \citep{Harris1984,Bridges1992}, with the most recent study estimating a total of $1610\pm30$ GCs and a corresponding specific frequency of $S_N=1.8\pm0.1$ \citep{Kang2022}. 
Another well-studied system is the nearby Sb-type spiral M81 \citep[e.g.,][]{Chies-Santos2022,Pan2022}.
Despite the challenges associated with identifying individual GCs, the estimate of the total number of GCs in M81 has remained relatively stable over time, with \citet{Perelmuter1995} quoting an estimated total of $N_\mathrm{GC} = 210 \pm 30$ and \citet{Nantais2011} finding 220--230 GCs, which yields $S_N\approx1.1$--1.2. 

The excellent image quality over a wide field-of-view offered by \Euclid\ 
\citep{EuclidSkyOverview}
holds greats promise for studying the outskirts of nearby galaxies, including their GC systems. It is in these regions that the signatures of hierarchical assembly histories are expected to be most apparent, and identification of GC candidates for further spectroscopic follow-up studies is therefore of considerable interest. 
Our main focus in this paper is on the old GC populations in and around the three galaxies, IC\,342, NGC\,2403, and Holmberg\,II. Hence, our lists of cluster candidates will be built primarily with GCs in mind, although they will also contain plenty of younger objects. 
Compared with previously published studies of GC systems based on the ERO data, such as Fornax \citep{EROFornaxGCs}, Dorado \citep{Urbano24}, and Perseus \citep{EROPerseusICL}, the Showcase galaxies are much closer. GCs are, therefore, expected to be better resolved, but also spread out over a larger area of the sky. Hence, besides identifying and characterising the GC systems of these three nearby galaxies and comparing with those of the Local Group counterparts, another aim of this paper is to gain a better understanding of how best to capitalise on the unique characteristics of \Euclid\ in this regime.
We leave aside the question of how exactly ``ancient'' GCs may or may not be related to the younger clusters observed in star-forming galaxies, and whether a physically meaningful distinction can even be made between these classes of objects. 
We will loosely (but perhaps not always entirely consistently) use the term GC to refer to objects with relatively red integrated colours, presumably indicating ``old'' ages ($\sim$ Gyrs). 

\section{Data}

The three galaxies were all observed with the same standard Reference Observing Sequence (ROS) used for the main Euclid Wide Survey \citep{Scaramella-EP1}, with four dithered subexposures yielding a total exposure time of $4\times566$\,s in the \IE\ filter of the VIS instrument \citep{EuclidSkyVIS} and $4\times87.2$\,s in each of the \YE, \JE, and \HE\ filters of the NISP instrument \citep{EuclidSkyNISP}. The pixel scale of the VIS instrument, $0\farcs1$\,pixel$^{-1}$ \citep{EuclidSkyVIS},
corresponds to a linear scale of 1.6--1.7\,pc at distances of 3.20--3.45\,Mpc, so that for a typical point-spread function (PSF) full width at half maximum (FWHM) of about 1.6\,pixel, 
the light profiles of star clusters with half-light radii of several pc are expected to be significantly broader than those of individual stars. 
The field-of-view is about $\ang{0.7;;}\times\ang{0.8;;}$ for both instruments, or about $39\,\mathrm{kpc}\times45\,\mathrm{kpc}$ at 3.2\,Mpc.
As described in \citetalias{ERONearbyGals} (where colour composite images can be found), the ERO data were processed independently of the standard \Euclid\ science ground segment pipelines, using a custom procedure described in detail in \citetalias{EROData}. 

For NGC\,2403 and IC\,342, the \Euclid\ imaging was supplemented with archival ground-based observations obtained with MegaCam on the Canada-France-Hawaii Telescope (CFHT). The MegaCam images cover a $\ang{1.04;;}\times\ang{1.15;;}$ field of view centred on each galaxy in the filters $u$, $g$, and $r$, as well as H$\alpha$ for IC\,342. The observations of NGC\,2403 were made on Jan 30, 2012, while those of IC\,342 were made between Jan 12 and Feb 11, 2021. 
The exposure times were $5\times60$\,s ($g$, $r$) and $5\times240$\,s ($u$) for NGC\,2403, and  
$7\times120$\,s ($g$), $21\times120$\,s ($r$), $7\times240$\,s ($u$), and $14\times360$\,s (H$\alpha$) for IC\,342. For each filter and galaxy, the MegaCam mosaics of 40 individual $2048\times4612\,\mathrm{pixel}$ CCDs were combined and co-added to a single image with a pixel scale of $0\farcs56$ since data were obtained in a FWHM seeing of typically $\sim1\farcs5$ for a diffuse emission motivated programme (down from the native camera sampling of $0\farcs19$). For the photometric calibration of the MegaCam data we used the zero-points specified in the image headers as derived by the CFHT \texttt{Elixir} pipeline \citep{Magnier2004}.

For Holmberg\,II we supplemented the \Euclid\ observations with deep imaging in the $g$ and $r$ bands from the Large Binocular Telescope (LBT). These images were acquired as part of the Smallest Scale of Hierarchy Survey \citep{Annibali2020}, using the Large Binocular Camera (LBC) which has a field of view of about $23\arcminute\times 23\arcminute$.
The images were acquired in binocular mode for a total 1\,h exposure time in each band, organised into 240\,s dithered exposures. During observations, the seeing was $\sim0\farcs8$. Image reduction and creation of the final $g$ and $r$ stacked mosaics were performed using a specific pipeline developed at INAF-OAR, as described in detail in \citet{Annibali2020}. 
The photometric calibration is based on the PanSTARRS1 \citep{Chambers2016,Flewelling2020} photometric catalog, after the PS1 photometry was transformed into the SDSS photometric system using the equations provided by \citet{Tonry2012}.

We note that the \Euclid\ and MegaCam/LBT photometry used in this paper is calibrated to the AB system, while literature $B$ and $V$ magnitudes are generally (approximately) Vega-based. 

\section{Overview of the three galaxies}
\label{sec:overview}

\begin{table}
\caption{Galaxy properties.}
\setlength{\tabcolsep}{3.25pt}
\smallskip
\label{tab:overview}
\smallskip
\small
\begin{adjustbox}{width=\columnwidth}
\begin{tabular}{lccccccc} \hline
& \\[-7pt]
 Galaxy & $D$ & $A_V$ & $B_T$ & $M_B$ & $M_V$ & $m_\mathrm{FIR}$ & SFR$_\mathrm{FIR}$\\
    & Mpc & mag & mag & mag & mag & mag & $M_\odot \, \mathrm{yr}^{-1}$ \\
    & & & \\ [-8pt]
\hline
 & & & \\ [-8pt]
IC\,342 & 3.45 & 1.53 & 9.1 & $-20.6$ & $-21.3$ & 6.95 & $1.7^{+1.7}_{-0.6}$ \\
NGC\,2403 & 3.2 & 0.11 & 8.93 & $-18.7$ & $-19.18$ & 8.63 & $0.35^{+0.35}_{-0.13}$ \\
Holmberg\,II & 3.32 & 0.087 & 11.1 & $-16.6$ & $-17.03$ & 12.84 & $0.0077^{+0.0077}_{-0.0029}$ \\
 & & & \\ [-8pt]
\hline
\end{tabular}
\end{adjustbox}
\tablefoot{Star-formation rates are computed from the far-infrared magnitudes $m_\mathrm{FIR}$ and distances $D$ using Eq.~(6) from \citet{Larsen2000}. 
}
\end{table}

Basic properties of the galaxies are listed in Table~\ref{tab:overview}. We assume the same distances and foreground extinctions as in \citetalias{ERONearbyGals}, while the integrated photometry is from the RC3 catalogue \citep{DeVaucouleurs1991} as provided through the NASA/IPAC Extragalactic Database (NED). The RC3 lists no $V$ magnitude for IC\,342, but we have assumed a colour of $(B-V)_0=0.7$, intermediate between those of $(B-V)_0=0.51$ for M33 and $(B-V)_0=0.87$ for M31. 
For Holmberg~II, we note that \citet{Bernard2012a} find a somewhat fainter integrated magnitude, $M_V=-16.3$.
Here and elsewhere, we follow the common practice of denoting reddening-corrected colours by the subscript '$0$'.
Star-formation rates, $\mathrm{SFR}_\mathrm{FIR}$, are computed from far-infrared (FIR) magnitudes ($m_\mathrm{FIR}$) and distances $D$ using Eq.~(6) from \citet{Larsen2000}. 

We correct the photometry for a single value of the foreground extinction, assumed to be uniform across the field. 
The $A_V$ values were converted to extinctions in each of the \Euclid\ filters, assuming a spectral template with $T_\mathrm{eff}=4500$\,K, $\log g = 2$, and $\mathrm{[Fe/H]}=-1$ (as an approximation to the spectrum of an old, moderately metal-poor GC), and the G23 extinction curve from \citet{Gordon2023}. 
The conversions are: $A(\IE)/A_V = 0.69$, $A(\YE)/A_V = 0.35$, $A(\JE)/A_V=0.23$, and $A(\HE)/A_V=0.15$. 
For the MegaCam filters the corresponding conversions are $A_u/A_V=1.50$, $A_g/A_V=1.14$, and $A_r/A_V=0.81$.
For the LBT photometry we used the Galactic foreground extinctions for SDSS $g$ and $r$ from \citet{Schlafly2011}, available through the NED. 
Although the extinction correction will depend somewhat on the spectral energy distribution of each source, especially for the very broad \IE\ filter, we have assumed a single correction for all sources. 

For a standard, approximately Gaussian GC luminosity function (in magnitude units), the turn-over is expected at $M_\mathrm{TO}(\IE)\approx-8$ \citep{EP-Voggel}.
For the distances and extinctions of NGC\,2403 and Holmberg\,II this implies an apparent magnitude of $m_\mathrm{TO}(\IE)\approx 19.7$, while the higher foreground extinction (and slightly greater distance) of IC\,342 places the expected turn-over at $m_\mathrm{TO}(\IE)\approx 20.9$. 

NGC\,2403 is an outlying member of the M81 group \citep{Tammann1968}. 
Its $M_B$ is about 0.3\,mag brighter than that of M33 and, 
like M33, it is a late-type spiral with no central bulge. Deep imaging of its outer parts has revealed an extended stellar component that can be traced to a  distance of 40\,kpc from the centre of the galaxy \citep{Barker2012} and
the galaxy also has a relatively undisturbed, symmetric and warp-free H{\sc i} disc \citep{DeBlok2008}. This points to NGC\,2403 having evolved in relative isolation until the present time, and now being on its first approach towards the M81 group \citep{Williams2013}, although there is evidence of an interaction with the dwarf galaxy DDO\,44 \citep{Carlin2019,Veronese2023}.
The SFR$_\mathrm{FIR}$ is lower by about a factor of two than the SFR of $\sim0.74 \, M_\odot \, \mathrm{yr}^{-1}$ of M33 \citep{Lazzarini2022}. However, \citet{Williams2013} found a SFR of $0.7 \, M_\odot \, \mathrm{yr}^{-1}$ for NGC\,2403. Hence, within the margins of uncertainty, the two galaxies may be considered to have similar SFRs. 
Prior to the \Euclid\ ERO observations, a number of studies had identified GC candidates in NGC\,2403
\citep{Tammann1968,Battistini1984,Drissen1999,Davidge2007,Forbes2022} and seven old GCs had been confirmed spectroscopically, with eight new candidates identified in \citetalias{ERONearbyGals}.
Based on the halo mass vs.\ GC number relation of \citet{Burkert2020}, \citet{Forbes2022} predicted a total population of 40--50 GCs, suggesting that a significant number of GCs may still await discovery. 

IC\,342 is one of the closest large spirals outside the Local Group, but its low Galactic latitude ($b=\ang{10.6;;}$) and considerable foreground extinction make it a challenging target for studies of stellar clusters, and old GCs especially. 
With $M_V\approx-21.3$, its luminosity is comparable to that of the Milky Way and somewhat lower than that of M31 (see the Introduction).
The SFR$_\mathrm{FIR}$ is similar to the current SFR of the Milky Way \citep[$2.0\pm0.7 \, M_\odot$~yr$^{-1}$;][]{Elia2022} and several times higher than that of M31 \citep[$\sim0.3$--$0.4 \, M_\odot$~yr$^{-1}$;][]{Tabatabaei2010}.
To our knowledge, there are no previous studies of GCs or other star clusters in IC\,342, other than the massive ($M \sim 6\times10^6 M_\odot$) nuclear star cluster \citep{Boker1999,Schinnerer2003}.
By comparison with the Milky Way and M31, one might expect to find 150--500 GCs in IC\,342. 

Finally, Holmberg\,II (UGC\,4305 = DDO\,50) 
has a luminosity similar to that of the SMC and we might expect to find a few several-Gyr old GCs in Holmberg\,II. 
Compared to SFR$_\mathrm{FIR}$, estimates from resolved colour-magnitude diagrams (CMDs) give a higher SFR of $\sim0.06 \, M_\odot \, \mathrm{yr}^{-1}$ averaged over the past 1\,Gyr \citep{Weisz2008}, but still slightly lower than the mean SFR of about $0.1 \, M_\odot \, \mathrm{yr}^{-1}$ over the past 2--3\,Gyr derived for the SMC with similar methods \citep{Harris2004}. Alternatively, we can compare the absolute FIR magnitudes of the SMC ($M_\mathrm{FIR}=-15.50$)
and Holmberg\,II ($M_\mathrm{FIR}=-15.13$) directly; the difference again suggests a slightly lower SFR for Holmberg\,II.
The star cluster population in Holmberg\,II was included in the HST-based study by \citet{Billett2002}, and the galaxy was also part of the ANGST \citep{Cook2012} and LEGUS \citep{Cook2019} surveys.

\section{Analysis}
\label{sec:analysis}

\subsection{Identification of cluster candidates}
\label{sec:identifying}

To search for star clusters, we first generated lists of sources in the VIS images using \texttt{SExtractor}  \citep{Bertin1996}.  We required a source to be detected as at least 6 connected pixels (\texttt{DETECT\_MINAREA=6}), each with a signal exceeding the mean background level by 8 standard deviations of the background noise (\texttt{DETECT\_THRESH=8}). 
An area of 6 pixels corresponds to a radius of 1.4\,pixels or $\approx2.3$\,pc at the  distances of our galaxies, and we therefore expect the light from a typical star cluster with a half-light radius of several pc to extend over a significantly larger area than 6\,pixels, so that this requirement should not significantly bias the detection against even fairly compact clusters. 
This expectation is borne out by the completeness tests described below (Sect.~\ref{sec:compl}).

We next converted the coordinates of sources detected in the VIS images to the NISP and MegaCam (LBT for Holmberg\,II) frames, using the World Coordinate System information in the image headers.
For the IC\,342 and Holmberg\,II images, no significant systematic offsets were found between the coordinate systems of the \Euclid\ and ground-based observations, whereas corrections of $0\farcs50$ in RA and $0\farcs36$ in Dec were applied to the MegaCam solution for NGC\,2403. We then used the \texttt{IRAF} version of \texttt{DAOPHOT} \citep{Stetson1987} to obtain aperture photometry on all images, using an aperture radius of $1\arcsec$ (10 pixels for the VIS images, 3.3\,pixels for NISP, 2\,pixels for MegaCam, and 4.4\,pixels for LBT). 
The transformed coordinates from the VIS images were used directly, with no recentering. 
For the \Euclid\ images aperture corrections of $-0.072$\,mag (\IE) and $-0.135$\,mag (\HE) were adopted from \citetalias{EROData}, as in \citetalias{ERONearbyGals}. For the ground-based images, aperture corrections from the $r=1\arcsec$ apertures were determined by measuring the differences between the magnitudes measured in those apertures and larger apertures for about 50 isolated stars in each frame. 
We assumed that these aperture corrections, determined for point sources, were also adequate for star clusters. 

\begin{figure}
\centering
\includegraphics[width=\hsize]{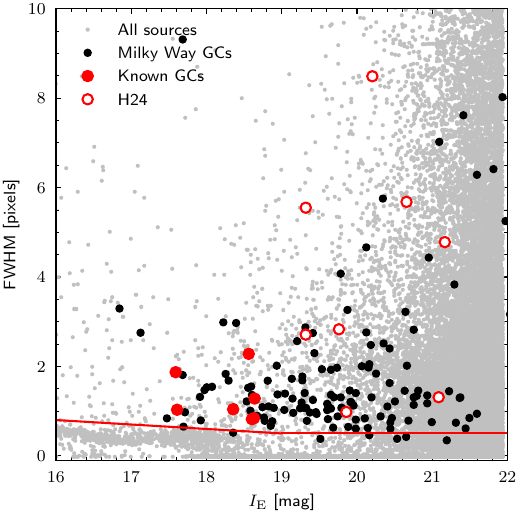}
\caption{PSF-corrected FWHM size as a function of \IE\ magnitude for sources in the NGC\,2403 field. Previously known GCs are shown with red filled circles while the new candidates identified in \citetalias{ERONearbyGals} are shown with open circles. Also included are data for Milky Way GCs, scaled to the distance of NGC\,2403 (black filled circles). The red line indicates our size cut for selection of cluster candidates. }
\label{fig:ccsel_n2403}
\end{figure}

The initial \texttt{SExtractor} runs detected more than 100\,000 sources in each field, of which only a very small fraction are star clusters. Hence, an additional selection based on the sizes and magnitudes of the sources was applied, followed by a visual inspection of the remaining candidates. 
We used the \texttt{ISHAPE} code \citep{Larsen1999a,Larsen2014} to measure PSF-corrected sizes for all sources brighter than $\IE = 22$ in each image. At this magnitude, the photometric uncertainties for the VIS images are generally less than 0.01\,mag, corresponding to a $\mathrm{S/N}>100$, ensuring accurate size measurements \citep{Larsen1999a}.
A PSF was first constructed for each image from about 100 individual, isolated stars, using the \texttt{PSF} task in \texttt{DAOPHOT}.
\texttt{ISHAPE} then measured the intrinsic sizes of the sources by convolving a chosen analytic model profile with the PSF and adjusting the FWHM of the model until the best match was obtained.  We assumed \citet{King1962}  profiles with a concentration index $c=r_\mathrm{t}/r_\mathrm{c}=30$ for tidal- and core radii $r_\mathrm{t}$ and $r_\mathrm{c}$.  
For these models, the conversion between the FWHM and half-light radius $r_\mathrm{h}$ is given by $r_\mathrm{h} = 1.48 \, \mathrm{FWHM}$ \citep{Larsen2014}. We note that, for a typical half-light radius of 3\,pc and distances of 3.20--3.45\,Mpc, 96--97\% of the total light of such a profile falls within our $r=1\arcsec$ photometric apertures.
We allowed \texttt{ISHAPE} to fit for both the major- and minor axes of the models; in the following we use the sizes measured along the major axis. 
In principle, the minor/major axis ratio of the model fit might be used as an additional selection criterion. However, while most Milky Way GCs appear fairly round in projection, star clusters in some galaxies, such as the LMC and NGC~6822, have a broader distribution of axis ratios, reaching values as low as $\sim0.6$ \citep{Goodwin1997,Huxor2013,Larsen2001c}. Therefore, to avoid introducing a bias against unusually elliptical objects, we have chosen not to use the axis ratio as a selection criterion here. A master catalogue containing the \Euclid\ VIS and NISP photometry, the \texttt{ISHAPE} size measurements, and the MegaCam/LBT ground-based photometry was then produced for each field. 

Cluster candidates were selected from the master catalogues by applying the same size cut as in \citetalias{ERONearbyGals}: $\mathrm{FWHM} > 0.5$\,pixels for $19 < \IE < 21.5$ and $\mathrm{FWHM}>0.5-(\IE-19)/10$\,pixels for $\IE < 19$. The size cut is indicated by the red line in Fig.~\ref{fig:ccsel_n2403}, which shows the \texttt{ISHAPE} PSF-corrected FWHM values as a function of the \IE\ magnitude for sources in the NGC\,2403 master catalogue. The previously known GCs are shown as filled red circles and the candidates from \citetalias{ERONearbyGals} are shown with open red circles. 
The figure also shows data for Milky Way GCs \citep{Harris1996}, scaled to a distance of 3.2\,Mpc and assuming a constant $V-\IE=0.5$. For the Milky Way GCs, the same conversion between FWHM and $r_\mathrm{h}$ given above has been assumed, although this depends somewhat on the concentration parameter and not all Milky Way GCs have exactly the same $c$ value. Nevertheless, it is clear that the vast majority of the Milky Way GCs would fall above the adopted size cut. 
A sequence of unresolved sources (stars) is visible at $\mathrm{FWHM}=0$\,pixels for magnitudes $\IE > 18$. At brighter magnitudes, point-like sources are saturated in the VIS images and \texttt{ISHAPE} thus tends to measure larger sizes, while more extended sources can still avoid saturation. 
The limit at FWHM = 0.5\,pixels corresponds to about 0.75\,pc at a distance of 3.2\,Mpc, or a half-light radius of about 1.1\,pc for the assumed King $c=30$ model profiles.

While the size cut eliminates most foreground stars, the remaining sources still include many non-clusters. Outside the main bodies of the galaxies, these are mainly background galaxies, but \texttt{SExtractor} also detects numerous sources in crowded regions of the disc and spiral arms, many of which remain in the candidate catalogue also after application of the size cut. It is difficult to ascertain the exact nature of many of these sources. Some are clearly regions of on-going or very recent star formation, associated with nebulosity, while others may be young star clusters or OB associations, or simply asterisms (chance alignments of a few relatively bright stars) or regions of higher than average surface brightness that happen to be picked up by \texttt{SExtractor}.
Background galaxies tend to be redder than GCs, but with some overlap, so that a colour cut can potentially help reduce the contamination, but a visual inspection was still found to be a necessary further step.

\subsection{Artificial cluster tests: expected appearance of GCs}
\label{sec:artclust}

\begin{figure}
\includegraphics[width=\hsize]{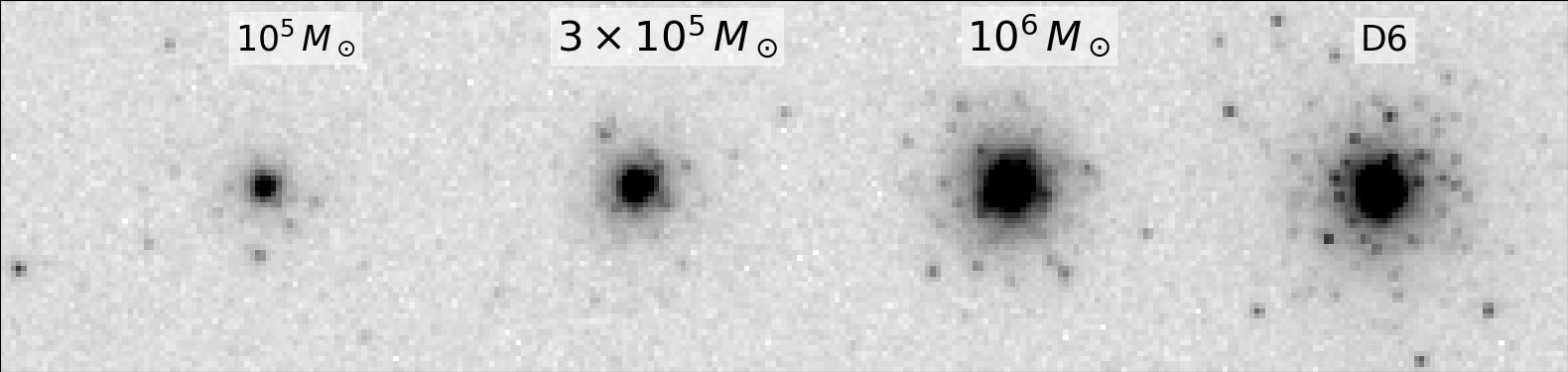}
\includegraphics[width=\hsize]{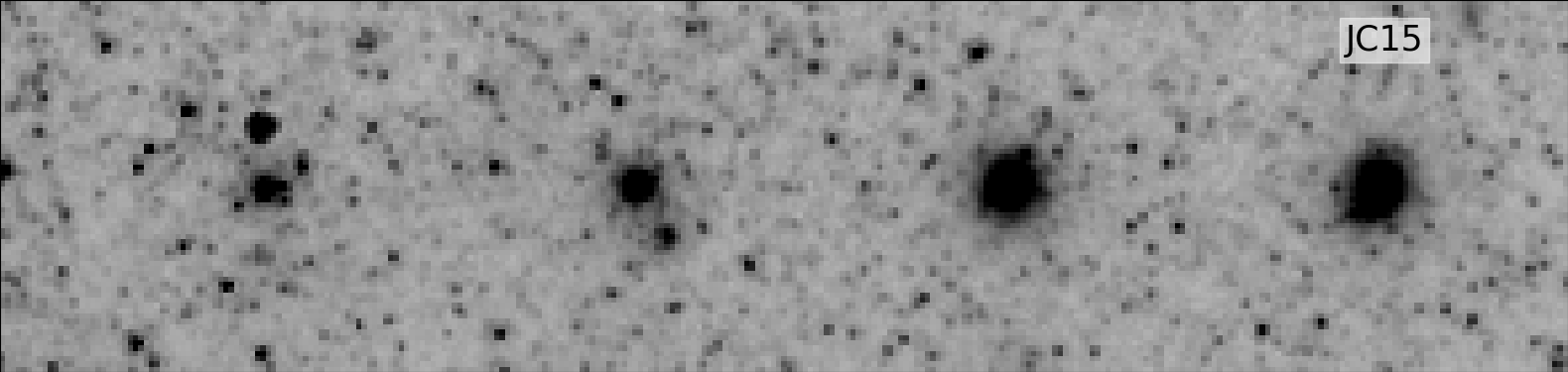}
\includegraphics[width=\hsize]{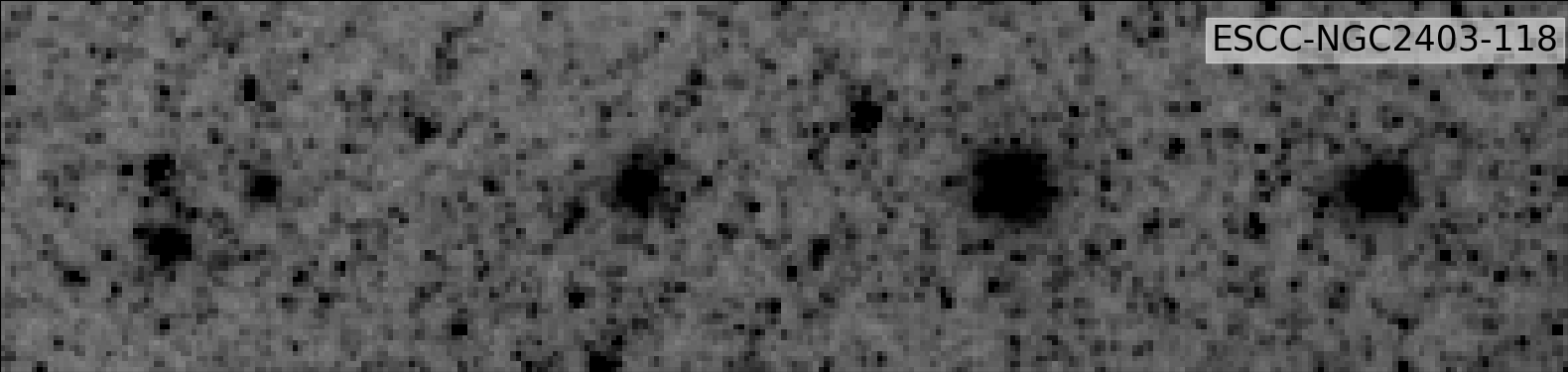}
\caption{\label{fig:simGCsN2403}Simulated images of GCs added to the \Euclid\ VIS image of NGC\,2403, next to the clusters D6 (top), JC15 (centre), and a new candidate closer to the centre of NGC\,2403 (bottom). In each panel, simulated GCs are shown for masses of $10^5 M_\odot$, $3\times10^5 M_\odot$, and $10^6 M_\odot$ (left to right)  a half-light radius of 3\,pc, and an assumed age of 10\,Gyr.
Each panel measures $295\times70$ VIS pixels ($29\farcs5\times7\farcs0$)}
\end{figure}

\begin{figure}
\includegraphics[width=\hsize]{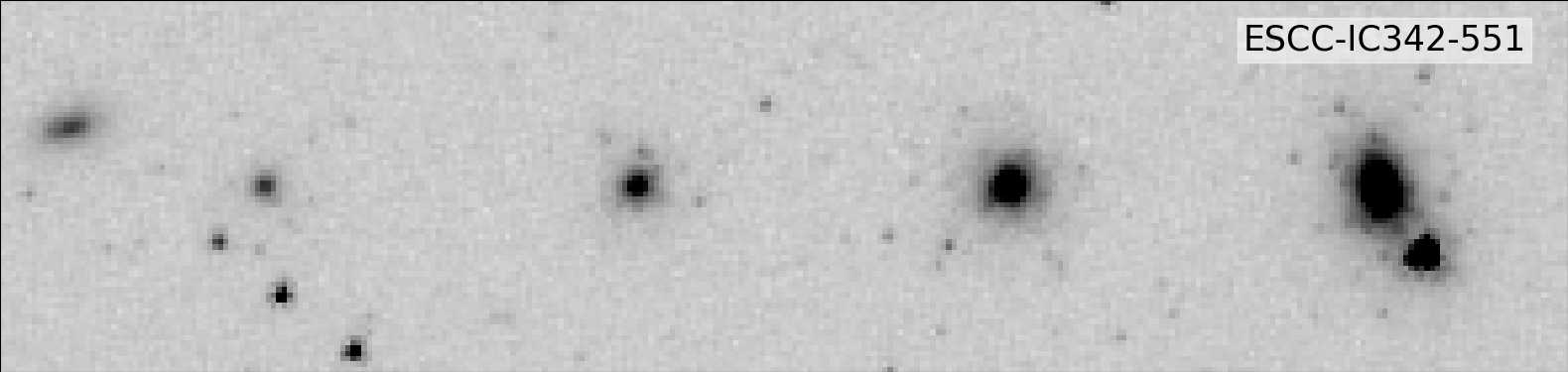}
\includegraphics[width=\hsize]{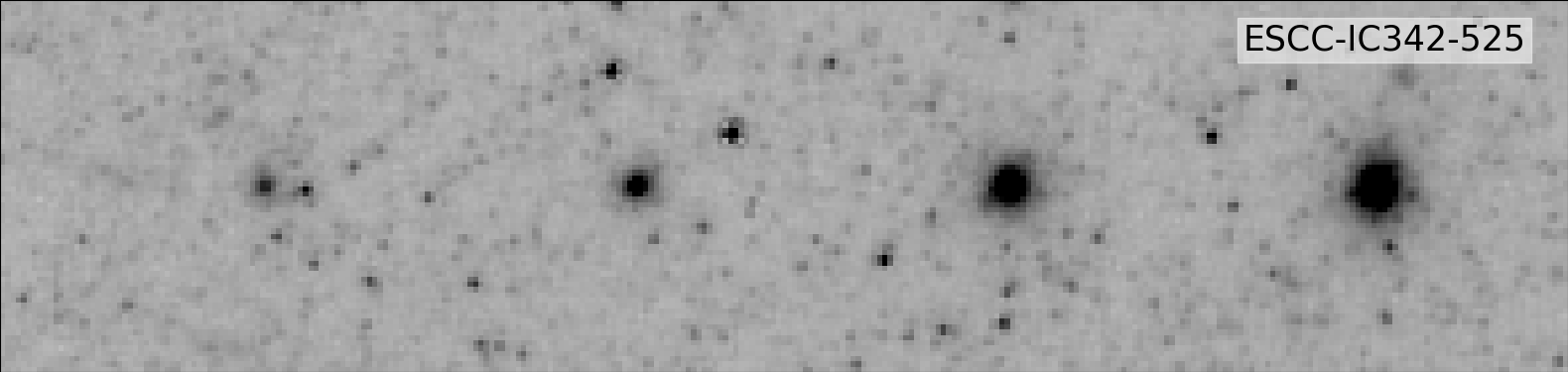}
\includegraphics[width=\hsize]{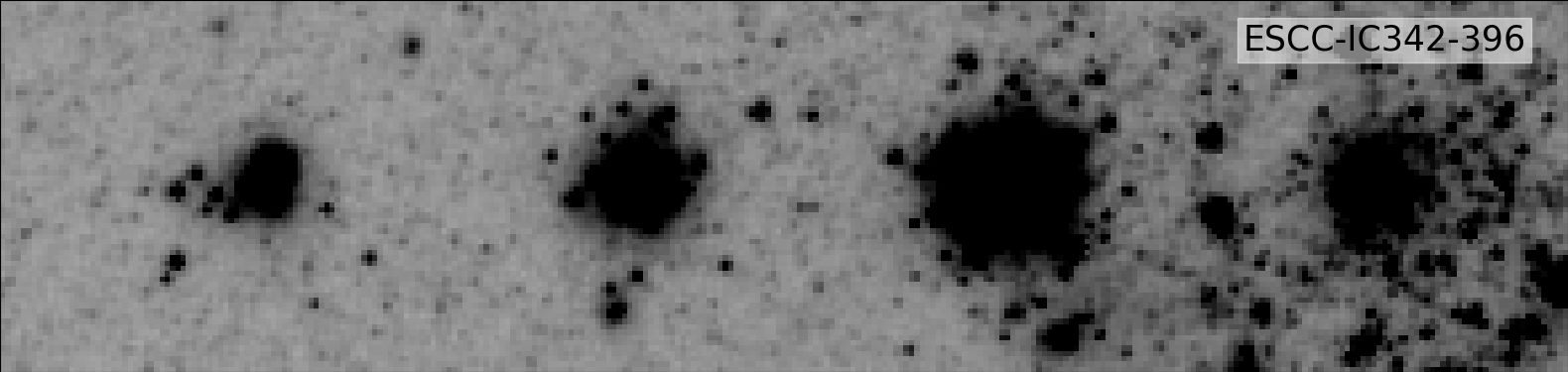}
\caption{\label{fig:simGCsI342}Top and centre panels: as Fig.~\ref{fig:simGCsN2403}, but for IC\,342. Bottom panel: simulated clusters with ages of 20\,Myr.}
\end{figure}

To assess how GCs are expected to look in the \Euclid\ images, we added artificial clusters to the VIS images of NGC\,2403 and IC\,342 at several different locations within the two galaxies. The artificial clusters were generated by sampling stellar masses at random from a \citet{Kroupa2001} IMF and distributing them spatially according to a probability distribution defined by a \citet{King1962} profile. \Euclid\ \IE\ magnitudes were then assigned to each star by interpolating in a PARSEC isochrone \citep{Bressan2012,Tang2014,Chen2015} with $\mathrm{[Fe/H]}=-1.0$ and an age of $t=10$\,Gyr. The absolute magnitudes from the isochrones were converted to apparent magnitudes in the \Euclid\ images by adding the distance modulus and $A(\IE)$ value appropriate for each galaxy.
The clusters were then simulated by adding these member stars to the VIS images with the \texttt{MKSYNTH} task in the \texttt{BAOLAB} package \citep{Larsen2014}, using the same PSFs as for the \texttt{ISHAPE} size measurements. 
Simulated clusters with masses of $10^5 M_\odot$, $3\times10^5 M_\odot$, and $10^6 M_\odot$ and a half-light radius of 3\,pc were added next to a known cluster (or cluster candidate), separated by $7\arcsec$.

Figure~\ref{fig:simGCsN2403} shows artificial clusters in NGC\,2403 added next to the known GCs D6 (at a projected separation of $14\farcm3$ from the centre of NGC\,2403), JC15 (at $4\farcm1$), and a new candidate nearer the centre (ESCC-NGC2403-118), at $1\farcm3$  
(we adopt the naming convention ESCC-Galaxy-ID, for Euclid Star Cluster Candidate, for the new star cluster candidates).
The three cases, which are all displayed with the same contrast settings, illustrate a range of crowding conditions encountered within the images.
Near D6, the field star density is very low and the more massive clusters are easily recognised as   resolved into individual stars in their outer regions. The median sky background in this field is about $\mu_{\IE} \approx 22.6$\,mag\,arcsec$^{-2}$, which is essentially the pure background sky level of the \Euclid\ \IE\ exposures for NGC\,2403 (\citetalias{ERONearbyGals}; \citealt{Scaramella-EP1}). However, even under these favourable conditions, confident identification of GCs with masses less than about $10^5 M_\odot$ based on (partial) resolution into individual stars is challenging. As crowding increases nearer the centre of the host galaxy, it becomes more difficult to identify clusters reliably. The field near JC15 has a median sky background of $\mu_{\IE} \approx 21.8$\,mag\,arcsec$^{-2}$, and the outer resolved regions of the GCs now start blending in with the general background of resolved field stars. 
In the inner field, with $\mu_{\IE} \approx 20.9$\,mag\,arcsec$^{-2}$, the outer resolved regions of the simulated clusters are now essentially impossible to distinguish, although the central parts are still spatially resolved, and the lowest-mass clusters appear visually similar to a number of other sources of uncertain nature visible in the field.  

Figure~\ref{fig:simGCsI342} shows artificial clusters added to the IC\,342 images next to sources ESCC-IC342-551 (at $15\farcm8$), ESCC-IC342-525 ($6\farcm6$), and ESCC-IC342-396 ($4\farcm2$). 
While the distances (and hence physical scales) in Fig.~\ref{fig:simGCsN2403} and \ref{fig:simGCsI342} differ only slightly, the increased extinction towards IC\,342 makes it more difficult to recognise individual RGB stars, and consequently also to use this as a robust criterion for identification of GCs.
Indeed, the tip-RGB at $M(\IE) \approx -3.3$ (according to the PARSEC isochrones) will appear at $\IE \approx 25.5$ in IC\,342 and at $\IE \approx 24.3$ in NGC\,2403. For IC\,342 this is about 0.7\,mag brighter than the expected $5\sigma$ detection limit for an isolated point source in the VIS images, while in NGC\,2403 (and Holmberg\,II) the brightest RGB stars are about 2\,mag brighter than the detection limit. 
Younger clusters, in which the brightest individual stars can be far more luminous than the tip-RGB, will be more easily identifiable. In the lower panel of Fig.~\ref{fig:simGCsI342} we have added artificial sources with the same masses and sizes as in the top panel, but with an age of 20\,Myr. The star cluster to the right in this panel is one of the brightest in the disc of IC\,342. The difference in brightness between a 10\,Gyr old population and a 20\,Myr population with the same mass is, of course, striking, as is the evident resolution into numerous bright stars in the younger object.  

\subsection{Artificial cluster tests: completeness and photometric uncertainties}
\label{sec:compl}

\begin{figure}
\centering
\includegraphics[width=\hsize]{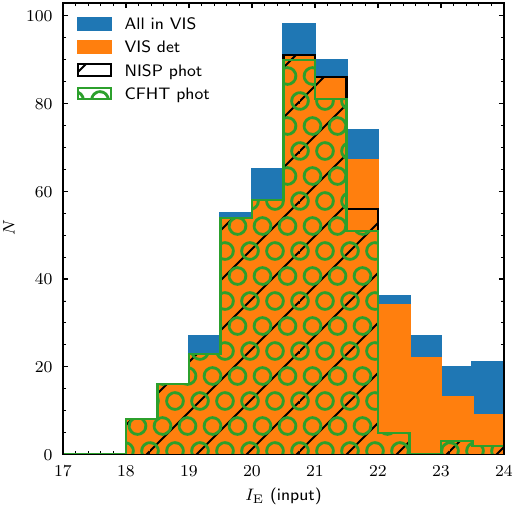}
\caption{Luminosity functions for simulated Milky Way GCs added to the \Euclid\ images of IC\,342. The distributions of input \IE\ magnitudes are shown for all clusters added within the \Euclid\ footprint (blue), those recovered in the VIS images (orange), and those that also have photometry in the NISP (black hatched) and MegaCam images (green circular-hatched).}
\label{fig:i342_mwgc_cmpl}
\end{figure}

\begin{figure}
\centering
\includegraphics[width=\hsize]{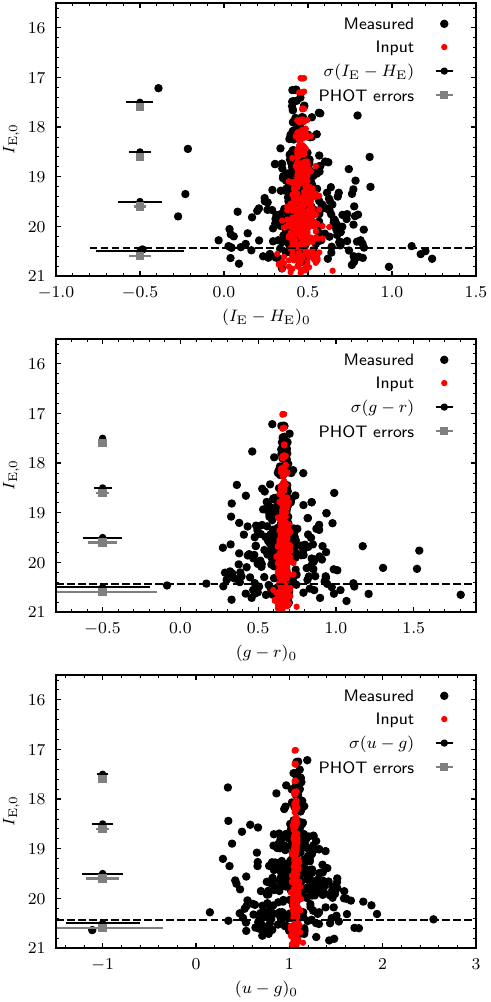}
\caption{Input- and measured colour-magnitude diagrams for artificial Milky Way GCs added to the IC\,342 images. The horizontal dashed lines indicate our magnitude limit for selection of cluster candidates.}
\label{fig:i342_mwgc_cmd}
\end{figure}

Tests of the type described in Sect.~\ref{sec:artclust} were also used to quantify the completeness of our candidate samples and the photometric uncertainties.
We restricted this analysis to IC\,342 as the most ``difficult'' case of the three galaxies, owing to its higher reddening and slightly greater distance compared to NGC\,2403 and Holmberg\,II. 

Of the 157 Milky Way GCs in the \citet{Harris1996} catalogue, 146 would fall within the footprint of the \Euclid\ VIS image if projected onto the disc of IC\,342. 
For the distance and extinction of IC\,342, our selection limit of $\IE=21.5$ for the visual inspection corresponds to $M(\IE) = -7.3$ or $M_V\approx-6.8$ for typical GC colours. 
Of the 146 Milky Way GCs falling within the VIS footprint, about 87 would have $\IE<21.5$ and are therefore bright enough to be included in our candidate list. 

To assess what fraction of a Milky Way-like GC population we would actually have identified in the \Euclid\ images, we converted the Galactic heliocentric ($X, Y$) coordinates of Milky Way GCs in the \citet{Harris1996} catalogue to corresponding CCD ($x, y$) coordinates in the \Euclid\ VIS image of IC\,342. For this conversion we assumed a distance of 8.4\,kpc from the Sun to the Galactic centre along the $X$-direction \citep{Ghez2008}.
We made no attempt to correct for any inclination of IC\,342.
For each cluster, we adopted the half-light radius and absolute visual magnitude, $M_V$, in the \citet{Harris1996} list and then added the clusters onto the \Euclid\ VIS and NISP images, as well as the MegaCam images, using the same simulation procedure described in Sect.~\ref{sec:artclust}.  
PSFs were constructed for each image in the same way as described for the VIS images (Sect.~\ref{sec:artclust}). 
For simplicity, we again assumed the same age (10\,Gyr) and metallicity ($\mathrm{[Fe/H]}=-1$) for all clusters. To improve the statistics, each cluster was added four times to the images by inverting the $X$- and $Y$-coordinates relative to the centre of IC\,342. In this way, a total of 584 artificial Milky Way GCs were added within the \Euclid\ footprint.
Figure~\ref{fig:ic342stamps_mwgc} shows VIS thumbnail cut-outs of a selection of the simulated Milky Way GCs brighter than $\IE=21.5$. Most of these stand out quite clearly above the background and would be confidently identified as non-stellar.

We repeated the \texttt{SExtractor} detection procedure, the \texttt{ISHAPE} size measurements, and the \texttt{DAOPHOT} photometry on the IC\,342 images with the artificial Milky Way GCs added to them. Figure~\ref{fig:i342_mwgc_cmpl} shows the distributions of input \IE\ magnitudes for all clusters added to the images, those recovered in the VIS images only, and those for which \texttt{DAOPHOT} was able to measure magnitudes in the NISP and MegaCam images.
The apparent sharp drop in clusters with NISP and MegaCam photometry below $\IE=22$ is due to the magnitude cut applied in the master catalogue. We note that this cut was applied to the measured magnitudes, not the input magnitudes, which explains the presence of a few objects with $\IE>22$. 
Of the 584 simulated clusters, 359 are brighter than our $\IE=21.5$ selection limit, and 336 (94\%) of these were recovered in the VIS image. 
For nearly all of the clusters detected in VIS and brighter than $\IE=21.5$, \texttt{DAOPHOT} was also able to measure magnitudes in the MegaCam and NISP images (330 and 336, respectively). 
A few clusters have measured sizes smaller than our adopted size cut of $\mathrm{FWHM}=0.5$~pixels, leaving 319 objects with VIS+MegaCam photometry after the size selection. Dividing by four, we would thus have detected about 80 out of the 87 Milky Way GCs contained within the \Euclid\ footprint that satisfy our magnitude limit. 
These numbers remain unchanged if we modify the \texttt{SExtractor} \texttt{DETECT\_MINAREA} parameter from 6 to 3 connected pixels.
Hence, down to our adopted selection limit, we expect to detect more than 90\% of any GCs, drawn from a population with properties similar to those of the Milky Way GC population, present within the \Euclid\ footprint.
These 80 detected clusters thus represent about half of the total population of 157 Milky Way GCs.

We can also use the artificial cluster tests to quantify the uncertainties on the photometry. 
The input- and measured MegaCam/\Euclid\ CMDs are shown in Fig.~\ref{fig:i342_mwgc_cmd} with
our (extinction-corrected) magnitude limit for selection of cluster candidates indicated by the horizontal dashed lines.
The scatter in the input colours is caused by the stochastic sampling of the stellar masses, which, like the photometric uncertainties, becomes more pronounced for fainter clusters. The stochastic sampling effects are also more significant in redder bandpasses, where the contribution from giants is more significant. 
The observed colours scatter fairly symmetrically with respect to the input values, with no evident bias as a function of magnitude. 
The black error bars show the computed dispersion of the measured colours (excluding outliers lying more than $3\sigma$ away from the mean), while the grey error bars show the mean uncertainties reported by \texttt{DAOPHOT}. 
It is clear that the formal photometric uncertainties underestimate the true uncertainties on the \Euclid\ GC colours, for which the stochastic sampling effects play an important role even at relatively bright magnitudes. 

\subsection{Visual inspection of the candidates}

\begin{figure}
\centering
\includegraphics[width=\hsize]{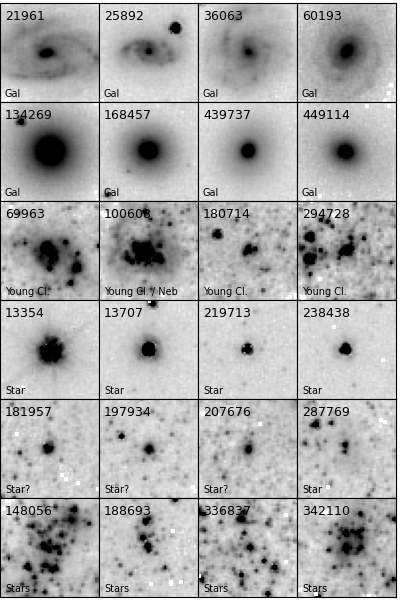}
\caption{Cut-out images of various sources in IC~342 that were not classified as class 1--4 cluster candidates. Each cut-out measures $5\arcsec\times5\arcsec$.}
\label{fig:ic342noncl}
\end{figure}

Guided by the simulations just described, the objects remaining after the size- and magnitude cuts were visually classified as ``Unlikely'', ``Maybe'' or ``Likely'' (globular) clusters. 
During this inspection, a few additional cluster candidates, not included in the master catalogues, were found
and manually added to our candidate lists. 
Objects that appeared clearly cluster-like, but whose detailed morphology was inconsistent with a classification as old GCs (such as those in the lower panel of Fig.~\ref{fig:simGCsI342}, resolved into stars much brighter than RGB stars, or objects clearly associated with on-going or very recent star formation), were classified as ``Young''.
The visual classification was done independently by three of us (AF, JH, SL) and our individual classifications were combined by a generalisation of the scheme described in Howell et al.\ (in prep.) to assign the following numerical classes to the objects: 
\begin{itemize}
    \item At least one ``Unlikely'' classification: class = 5;
    \item else: at least one ``Young'' classification: class = 10;
    \item else: sources classified as ``Likely'' by all: class = 1;
    \item else: sources with at least as many ``Likely'' as ``Maybe'' classifications: class = 2;
    \item else: sources with at least one ``Likely'': class = 3;
    \item else: (sources classified as ``Maybe'' by all): class = 4.
\end{itemize}
Objects in the ``Young'' (class=10) category will be treated separately throughout the remainder of the paper. 

A selection of sources in IC\,342 that passed the first round of size/magnitude criteria but were not included in the list as class 1--4 cluster candidates are shown in Fig.~\ref{fig:ic342noncl}. Since most of these do not have an ESCC (\Euclid\ star cluster candidate) identifier, we identify them by the IDs in our master catalogue. 
Some of these are very obvious late-type galaxies (first row), while others are almost certainly early-type galaxies (second row) with very smooth profiles that show no hint at all of resolution into individual stars, despite being located in regions of low background far from the main body of IC\,342. Clearly, such cases can be more ambiguous when projected against the disc of our target galaxies. The third row shows four objects in the ``Young'' category. 
In addition, some saturated foreground stars remained in the list even after the size cut but were generally easy to recognise by their sharp edges and diffraction spikes (fourth row). Some such cases may also be recognisable due to their non-zero \textit{Gaia} proper motions and/or parallaxes, 
but this will not always be the case; for example, the relatively faint object \#238438 ($\IE=20.56$) has proper motions of
$\mathrm{pm}_\mathrm{RA}=+0.25\pm0.30$\,mas\,yr$^{-1}$,
$\mathrm{pm}_\mathrm{DE}=-0.51\pm0.45$\,mas\,yr$^{-1}$, and a parallax of $p = -1.27\pm0.52$\,mas, and would not be easily identifiable as a foreground star based on \textit{Gaia} DR3 astrometry, while object \#219713 ($\IE=20.55$) has no \textit{Gaia} DR3 astrometry at all.
Other sources, such as those in the fifth row, may also be individual luminous stars in IC\,342 (or foreground stars projected against the disc), but could also be compact clusters.  The final row  with objects labelled ``Stars'' illustrates the diversity of morphologies seen in crowded regions that frequently exhibit some level of ``clustering'', but have not generally been included in our lists of cluster candidates. 

\begin{table}
\caption{Statistics for cluster candidate selection.}
\setlength{\tabcolsep}{3.25pt}
\smallskip
\label{tab:ccsel}
\smallskip
\small
\begin{adjustbox}{width=\columnwidth}
\begin{tabular}{lcccccccc} \hline
 & \\[-7pt]
 Galaxy & $N_\mathrm{Master}$ & $N_\mathrm{Sel}$ & $N_\mathrm{C=10}$ & $N_\mathrm{C=1}$ & $N_\mathrm{C=2-4}$ & $ N_\mathrm{IAC}$ & $N_\mathrm{GCC}$ & $N_\mathrm{Unknown}$\\
 & & \\[-8pt]
\hline
& & & \\[-8pt]
IC\,342 & 51815 & 6835 & 246 & 39 & 273 & 172 & 111 & 29 \\
NGC\,2403 & 47544 & 9073 & 72 & 43 & 105 & 81 & 50 & 17 \\
Holmberg\,II & 12475 & 3254 & 7 & 7 & 6 & 6 & 7 & 0 \\
\hline
\end{tabular}
\end{adjustbox}
\tablefoot{The column $N_\mathrm{Young}$ lists the number of objects visually identified as ``young'' cluster candidates. The columns $N_\mathrm{Likely}$ and $N_\mathrm{Maybe}$ give the numbers of potential GCs visually identified as ``Likely'' or ``Maybe'' cluster candidates. Of the clusters in the $N_\mathrm{Likely}$ and $N_\mathrm{Maybe}$ categories, $N_\mathrm{GCC}$ indicates the number of GC candidates remaining after colour selection, $N_\mathrm{IAC}$ the intermediate-age candidates, and $N_\mathrm{Unknown}$ objects with missing/ambiguous colours. 
}
\end{table}

Table~\ref{tab:ccsel} lists the number of sources remaining after each stage of the selection: $N_\mathrm{Master}$ is the total number of sources in each master catalogue,  $N_\mathrm{Sel}$ the number of objects remaining after the magnitude and size selection, and $N_\mathrm{C=10}$, $N_\mathrm{C=1}$ and $N_\mathrm{C=2-4}$ the number of young (class 10), class 1, and class 2--4  candidates. 
For NGC\,2403, the class 1--4 candidates listed include the seven previously known old GCs and the eight new candidates identified in \citetalias{ERONearbyGals}.
Of the 13 class 1--4 candidates in Holmberg\,II, 9 are included in previous work (Fig.~\ref{fig:HoIIstamps1}).
Of the three cluster candidates identified by \citet{Billett2002}, we classify two as background galaxies (BHE-1, BHE-2), while BHE-3 (our ESCC-HoII-012) is in common with the LEGUS sample (\citealt{Cook2019}; their \#11). We have an additional five candidates in common with LEGUS and three clusters in common with ANGST \citep{Cook2012}.
The remaining columns in Table~\ref{tab:ccsel}, $N_\mathrm{IAC}$, $N_\mathrm{GCC}$, and $N_\mathrm{Unknown}$ give the classification statistics for the class 1--4 clusters as intermediate-age or GC candidates based on the MegaCam/LBT photometry, as will be explained and discussed in more detail below. 

\subsection{Photometry of the cluster candidates}

\begin{figure}
\centering
\includegraphics[width=\hsize]{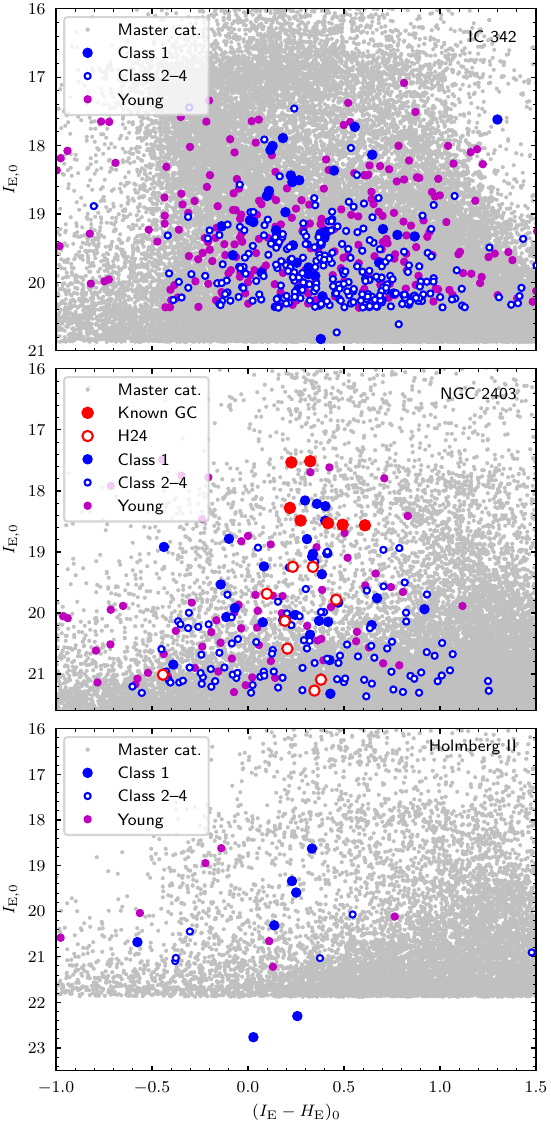}
\caption{\Euclid\ CMDs. The gray dots show all sources in the master catalogue for each galaxy, while the open and filled blue circles are our new candidates classified as ``possible'' or ``likely'', respectively. For NGC~2403, the previously known GCs are shown in red. Formal uncertainties on the photometry are generally less than 0.01\,mag, but see Sect.~\ref{sec:compl} for a detailed discussion of the uncertainties.}
\label{fig:CMDs_IH_I}
\end{figure}

\begin{figure}
\centering
\includegraphics[width=\hsize]{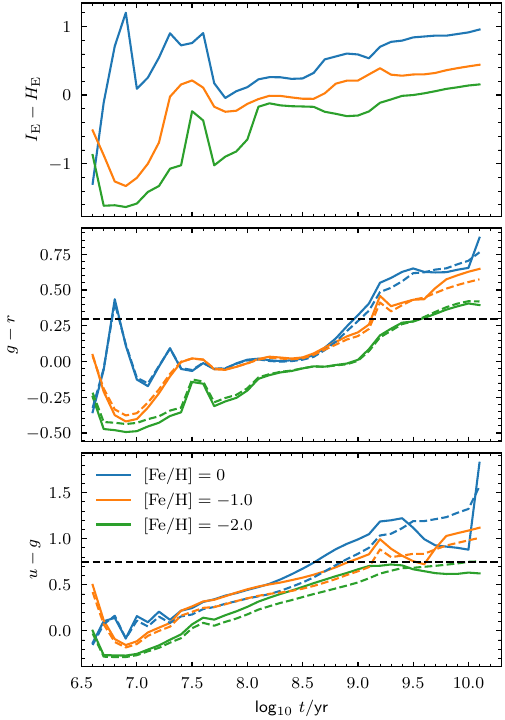}
\caption{PARSEC simple stellar population model colours versus age. Top: \Euclid\ $\IE-\HE$. Middle:  MegaCam $g-r$. Bottom: MegaCam $u-g$. For the MegaCam filters, dashed and solid lines indicate models for the old and new MegaCam filters, respectively. The horizontal dashed lines indicate the colour criteria for selection of old GC candidates.}
\label{fig:SSPcol_age}
\end{figure}

\begin{figure}
\centering
\includegraphics[width=\hsize]{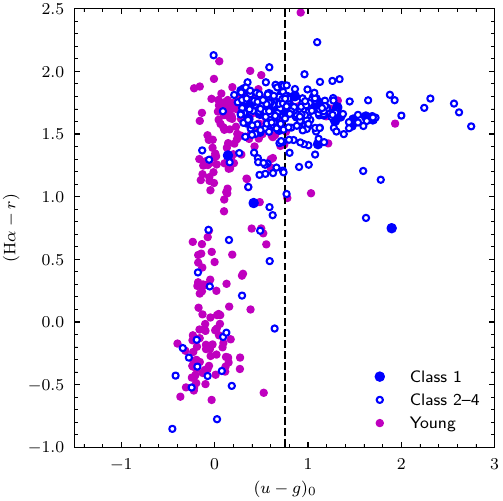}
\caption{MegaCam $\mathrm{H}\alpha-r$ vs.\ $(u-g)_0$ two-colour diagram for sources in IC\,342. Symbols as in Fig.~\ref{fig:CMDs_IH_I}. The dashed line indicates our colour cut for selection of old GC candidates.}
\label{fig:TCDs_ug_rH}
\end{figure}

\begin{figure}
\centering
\includegraphics[width=\hsize]{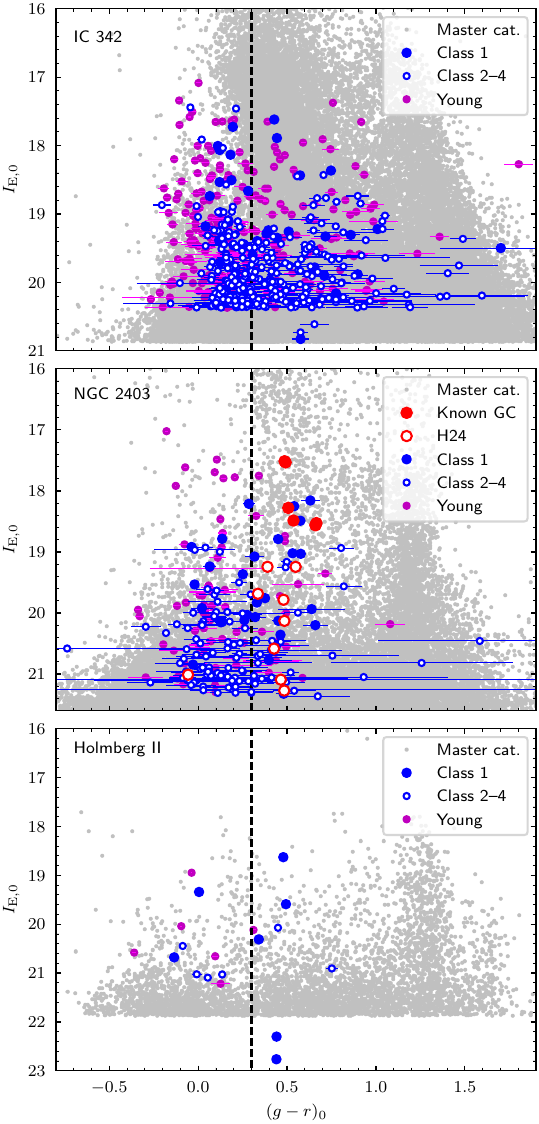}
\caption{\Euclid\ \IE\ vs. MegaCam $(g-r)_0$ CMDs. For Holmberg\,II, the $g-r$ photometry is from LBT.
Symbols are the same as in Fig.~\ref{fig:CMDs_IH_I}.}
\label{fig:CMDs_gr_I}
\end{figure}

In Fig.~\ref{fig:CMDs_IH_I} we show the \Euclid\ $(\IE-\HE, \IE)_0$ CMDs for the cluster candidates in each galaxy, together with all sources in the master catalogues. The best cluster candidates (class 1, i.e, classified as ``Likely'' by all of us) are indicated with filled blue circles and the more uncertain candidates (class 2--4, i.e, with at least one ``Maybe'' classification) with open blue circles.  Class 10 (young) candidates are marked with smaller magenta symbols.
For NGC\,2403, we also indicate the GCs identified prior to the \Euclid\ ERO observations (solid red circles) and those identified in \citetalias{ERONearbyGals} (open red circles). 
Compared to NGC\,2403, IC\,342 hosts a higher proportion of relatively blue, luminous clusters, which may suggest differences in the age- and/or metallicity distributions of the cluster populations. However, there is also a great deal of overlap between the colours of the class 10 candidates and the class 1--4 candidates, suggesting that the \Euclid\ colours by themselves may not be very effective at discriminating between clusters of different ages. 
In Holmberg\,II, the CMD is clearly much more sparsely populated with cluster candidates than in the two larger galaxies. Two very faint cluster candidates, but nevertheless classified as class 1, were noticed during the visual inspection and added manually to the master catalogue.
A few manually added faint candidates in IC\,342 are also visible in the CMD.

\subsubsection{Simple stellar population model colours}
\label{sec:sspcol}

To better understand how the various colours depend on age and metallicity, we plot simple stellar population (SSP) model colours in Fig.~\ref{fig:SSPcol_age}. The colours are shown as a function of age for three different metallicities ($\mbox{[Fe/H]} = -2$, $-1$, and $0$) for models based on PARSEC isochrones. It is clear that the \Euclid\ $\IE-\HE$ colours (top panel) alone are insufficient for determining whether a source might be an old, relatively metal-poor GC or a younger, more metal-rich star cluster: for example, a 10\,Gyr old cluster with $\mathrm{[Fe/H]}=-2$ is predicted to have an $\IE-\HE$ colour similar to that of a Solar-metallicity cluster with an age of around 100\,Myr. Both combinations are quite likely to be found in a giant spiral like IC\,342. While it was found in \citetalias{ERONearbyGals} that the old GCs in NGC\,2403 do have redder colours than the young objects, the models in the upper panel of Fig.~\ref{fig:SSPcol_age} suggest that the \Euclid\ colours in general are not ideal for discriminating between young and older star clusters. 

The two lower panels in the figure show models for the CFHT MegaCam colours.  Dashed and solid  lines indicate models for the old and new MegaCam filters, for observations made before and after 2015, respectively. We note that the NGC\,2403 observations were made with the old filters, and those of IC\,342 with the new ones. 
The $ugr$ colours provide better age discrimination than the \Euclid\ colours, although a significant age-metallicity degeneracy is still present at ages older than about 1\,Gyr. Adding to this fundamental uncertainty is the unknown contribution from dust reddening internal to the galaxies. Ultimately, these degeneracies can only be alleviated by means of spectroscopic observations and/or resolved CMDs of the clusters. Nevertheless, a selection based on the $u-g$ and $g-r$ colours, with
$(u-g)_0>0.75$ and $(g-r)_0>0.30$, as indicated by the horizontal dashed lines in Fig.~\ref{fig:SSPcol_age}, should allow us to provide a first, tentative list of clusters older than about 1~Gyr, and we will adopt these criteria for selection of ``old'' GC candidates throughout the remainder of this paper. 

\subsubsection{The effect of line emission}

The $r$-band filter includes the H$\alpha$ line, and can thus be affected by line emission, which is not included in the PARSEC SSP models.  The H$\alpha$ imaging of IC\,342 can help us quantify the effect. To this end, Fig.~\ref{fig:TCDs_ug_rH} shows a H$\alpha-r$ vs.\ $u-g$ two-colour diagram for sources in IC\,342.
Photometry was carried out on the H$\alpha$ images in the same way as for the $ugr$ images. 
Many of the young (class 10) candidates have a strongly enhanced flux in the H$\alpha$ images (i.e., a lower value of H$\alpha-r$), confirming the visual classification of these sources as young. In contrast, none of the sources with $u-g$ colours redder than our adopted selection limit for old GC candidates, indicated by the vertical dashed line, show significantly enhanced H$\alpha$ emission. We therefore expect that a selection based on the combination of $(u-g)_0$ and $(g-r)_0$ colours will indeed lead to a reasonably clean list of candidate old clusters.

\subsection{Globular cluster candidate selection: MegaCam and LBT photometry}
\label{sec:megalbt}

\begin{figure}
\centering
\includegraphics[width=\hsize]{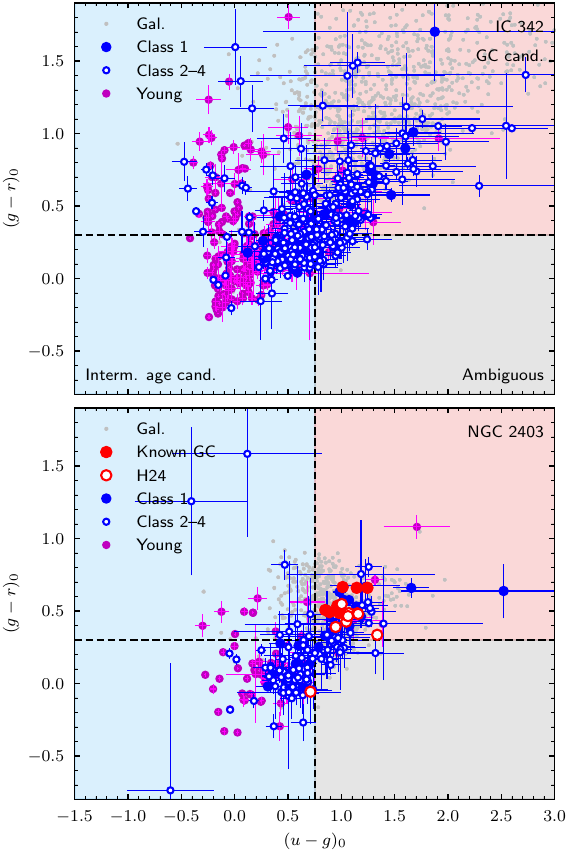}
\caption{MegaCam-only two-colour diagrams. The coloured quadrants indicate the classification of our class 1--4 candidates as GC candidates (red), intermediate-age candidates (blue), or ambiguous (grey). Small grey points indicate sources classified as galaxies during the visual inspection.}
\label{fig:TCDs_ug_gr}
\end{figure}

In  
Fig.~\ref{fig:CMDs_gr_I} we show the MegaCam/LBT $(g-r)_0$ vs.\ \Euclid\ \IE\ CMDs with vertical dashed lines indicating our adopted colour cut for selection of old GC candidates. 
We see that all of the known GCs in NGC\,2403, as well as most of the candidates from \citetalias{ERONearbyGals}, fall to the right of the colour cut, consistent with these objects being old GCs.  Again, many of the young sources have colours overlapping with those expected for old GCs owing to the effect of line emission. However, for the purpose of selecting old GCs, this ambiguity is removed by the inclusion of the $u-g$ colour, except for Holmberg\,II (where, however, the visual inspection will still help us identify the youngest sources).
The lack of stars brighter than $\IE\approx18-19$ in Holmberg\,II is caused by saturation in the deep LBT exposures.

Figure~\ref{fig:TCDs_ug_gr} shows a two-colour diagram of $(g-r)_0$ vs.\ $(u-g)_0$. 
Based on this diagram, we will henceforth refer to class 1--4 objects in the upper right quadrant of this figure as \emph{GC candidates} (GCC), those in the two left quadrants as \emph{intermediate-age candidates} (IAC), and those in the lower right quadrant as \emph{ambiguous}. We reserve the term \emph{young} cluster candidates for our class 10 sources.
For Holmberg\,II, this categorization is based on the $g-r$ colours only.
The numbers of objects in each of these categories are listed in the last three columns of Table~\ref{tab:ccsel}. Very roughly, the candidates in the young category probably have ages up to about 10\,Myr, the IAC sources up to $\sim1$\,Gyr, and GCC sources ages greater than about 1\,Gyr. We emphasise that these age ranges are only rough indications. 
Thumbnail cut-outs from the VIS images of the class 1--4 candidates are shown in Fig.~\ref{fig:ic342stamps1yc}-Fig.~\ref{fig:HoIIstamps1}, with the class indicated in each cut-out. 

Based on the MegaCam colour selection, we identify 50 GC candidates in NGC\,2403 and 111 in IC\,342 (Table~\ref{tab:ccsel}). Based on the LBT $g-r$ colours alone, 7 GC candidates are identified in Holmberg\,II.
The status of these sources as GC \emph{candidates} should be strongly emphasised, but we note that the total number of GC candidates in NGC\,2403 is quite similar to that expected based on the total mass of the galaxy. The 111 candidates in IC\,342 may be compared with the 157 known GCs in the Milky Way and more than 450 in M31, where we recall from Sect.~\ref{sec:compl} that we would have expected to detect about 80 of the 157 Milky Way GCs at the distance of IC\,342.
For comparison with Holmberg\,II, we note that the SMC, of comparable luminosity, hosts only one object that is traditionally classified as an old GC, NGC\,121, although it hosts several clusters with ages $\gtrsim1$\,Gyr \citep{Glatt2009,Parisi2014}. 

\subsection{Spatial distributions}
\label{sec:spatial}

The spatial distributions of the class 1--4 cluster candidates are plotted on top of the VIS images in Figs.~\ref{fig:ic342cc_yc_gc}--\ref{fig:HoIIcc_yc_gc}, colour-coded according to their classification as intermediate-age (blue), old (red), or ambiguous (black circles) based on the MegaCam colours.
The intensity scale is logarithmic in order to emphasise the faint outer parts of the galaxies. 
We do not include the class 10 sources in these figures. 

Many of the IAC sources in IC\,342 tend to be aligned with structure in the underlying disc, such as the spiral arms, while the old GC candidates are somewhat more uniformly distributed. Nevertheless, some GC candidates also align with disc structure, such as the outer spiral arm extending towards the south-west (also compare with the maps in \citetalias{ERONearbyGals}). Keeping in mind the relatively crude age dating based on the MegaCam colours, it is possible that some of these are somewhat younger disc objects rather than ancient GCs associated with the spheroidal component(s) of IC\,342. 
The median projected galactocentric distances of the IAC and GCC sources are 5.3\,kpc and 5.7\,kpc, respectively. For comparison, the median Galactocentric distance of the Milky Way GCs in the \citet{Harris1996} catalogue, projected onto the Galactic plane, is 5.0\,kpc. For the M31 GC system, we add the PAndAS clusters identified by \citet{Huxor2014} to the confirmed GCs in v.5 of the Revised Bologna Catalogue \citep{Galleti2004}, where the latter already includes the PAndAS clusters from \citet{Huxor2008}. This yields a median projected galactocentric distance of confirmed M31 GCs of 7.0\,kpc. 

For NGC\,2403 (Fig.~\ref{fig:ngc2403cc_yc_gc}), the spatial distribution of the GCC sources is noticeably more extended (median distance = 3.7\,kpc) than that of the IAC sources (2.7\,kpc). There is a deficit of sources near the centre, in part because a significant number of sources are missed by \texttt{SExtractor}, probably due to the difficulty of deblending detections in the crowded inner regions. 
Apart from the cluster ESCC4, all of the IAC sources are located within $7\arcmin$ of the centre of NGC\,2403, while 12 old GC candidates are found outside this radius. 
It is also interesting to note that the distribution of the old GCs appears to be aligned with the major axis of NGC\,2403, except for the cluster ESCC5 (towards the south). This might suggest that most of the GCs in NGC\,2403 follow a more disc-like distribution. As in IC\,342, it is possible that some of these objects are younger clusters with ages of only a few Gyr. The impression of a disc-like distribution may be partly driven by the four clusters located to the north-west, roughly aligned along an extension of the major axis. An alternative interpretation is that these GC candidates might be associated with the stream connecting DDO\,44 and NGC\,2403 \citep{Veronese2023,Carlin2024}, extending northwards from the western side of NGC\,2403.

For Holmberg\,II the distribution of our new cluster candidates is noticeably asymmetric and skewed towards the western side of the galaxy. Figure~\ref{fig:HoIIcc_yc_gc} also includes our class 10 sources and candidates from the literature and it can be seen that these are more evenly distributed, perhaps even with an opposite asymmetry.  The asymmetry may reflect an age gradient across the galaxy, with more active/recent star formation on the eastern side, where the most prominent H{\sc ii} regions are also found. Interestingly, the distribution of H{\sc i} gas exhibits a similar asymmetry, being compressed on the south-eastern side and with a more diffuse extension in the opposite direction \citep{Bureau2002}. 
The two faintest objects in our list, ESCC-HoII-001 and ESCC-HoII-002, which were manually added, are not included in any of the existing catalogues, although ESCC-HoII-001 is contained within the field-of-view of a single parallel HST/WFC3 F775W observation with an exposure time of 490\,s (programme ID 16359). The HST observation shows a faint source at the position determined from the \Euclid\ image, contaminated by several cosmic ray hits. We discuss these two outer sources in more detail below (Sect.~\ref{sec:gcc_cm}). 

\subsection{Comparison with HST imaging}

\begin{figure}
\centering
\includegraphics[width=\hsize]{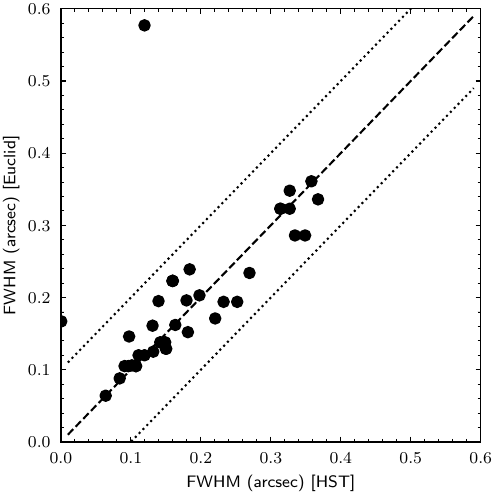}
\caption{Comparison of sizes for GC candidates in IC\,342 measured on the \Euclid\ VIS images and on HST/ACS F606W images.}
\label{fig:fwhm_hst_euclid}
\end{figure}

On the VIS images of IC\,342 and NGC\,2403 we have indicated the coverage of existing HST/ACS observations in the F435W/F606W filters. The enormous gain in coverage by the \Euclid\ observations is immediately visible. Also included for NGC\,2403 is the JWST/NIRCAM pointing of Programme ID 1638, which contains none of our candidates, but illustrates the difference in FOV sizes of \Euclid\ vs.\ JWST. 

Within the HST footprints for IC\,342, 41 of our 111 GC candidates and 69 of our 172 IAC candidates are contained. The HST/ACS F606W cut-outs of our GC candidates in IC\,342 are shown in Fig.~\ref{fig:hst_ic342stamps1gc}.
Based on inspection of the HST images of these candidates, our classifications based on the \Euclid\ images did not change significantly for all but a few sources. 
As a check of the size measurements on the \Euclid\ images we also measured sizes for the GC candidates contained within the HST images, again using \texttt{ISHAPE} with custom-made PSFs constructed from isolated stars in the HST images. 
The comparison is shown in Fig.~\ref{fig:fwhm_hst_euclid}. Most of the measurements lie well within $\pm0\farcs1$ (i.e.\ one \Euclid\ VIS pixel) of the 1:1 relation, as indicated by the dashed and dotted lines in the figure. One source, ESCC-IC342-241, is unresolved in the HST images and is thus a probable star, whereas the fits to the \Euclid\ images returned a PSF-corrected FWHM of 1.67\,pixels. 
The reason for the discrepancy is possibly related to the fact that the source is located only $60\arcsec$ from the nucleus of IC\,342, where the background is very high.
Excluding this and another outlier falling outside the dotted ($\pm0\farcs1$) lines in Fig.~\ref{fig:fwhm_hst_euclid}, the dispersion of the FWHM measurements around the 1:1 relation is $0\farcs033$ (1/3 of a VIS pixel), corresponding to a dispersion of 0.82\,pc on the half-light radii at the distance of IC\,342. We note that this is about 4 times larger than the uncertainty quoted by \citet{EP-Voggel} for GCs in Fornax. While the scatter here includes contributions from both the \Euclid\ and HST size measurements, 
this underscores the need to assess the uncertainties on this type of analysis individually for each case. 

\subsection{Sizes of the cluster candidates}
\label{sec:sizes}
\begin{figure}
\centering
\includegraphics[width=\hsize]{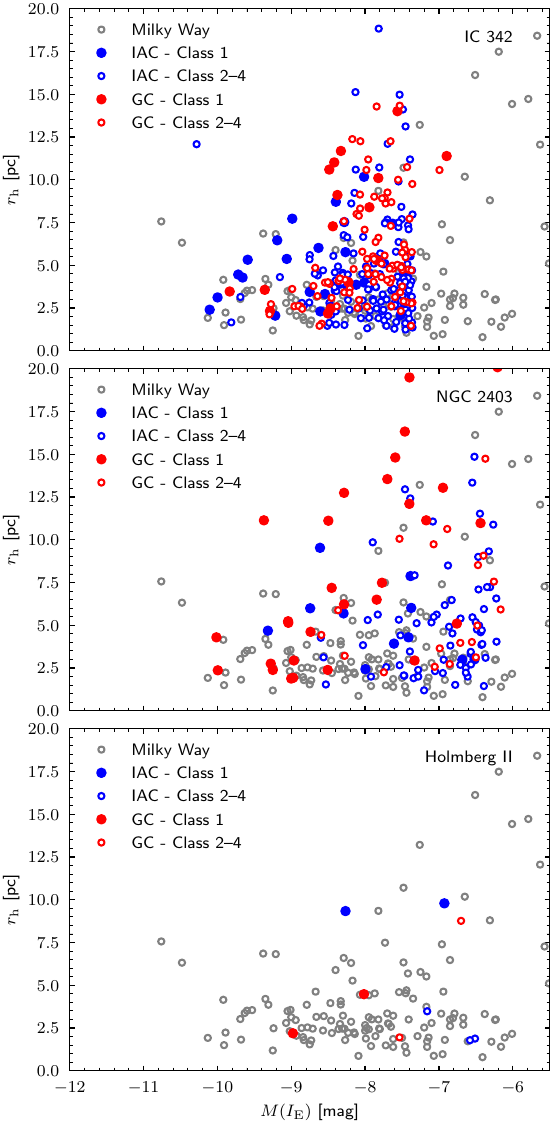}
\caption{Half-light radius as a function of  $M(\IE)$ for star cluster candidates. In each panel, the corresponding data for Milky Way GCs are shown for reference.}
\label{fig:sizes_gc}
\end{figure}

\begin{figure}
\centering
\includegraphics[width=\hsize]{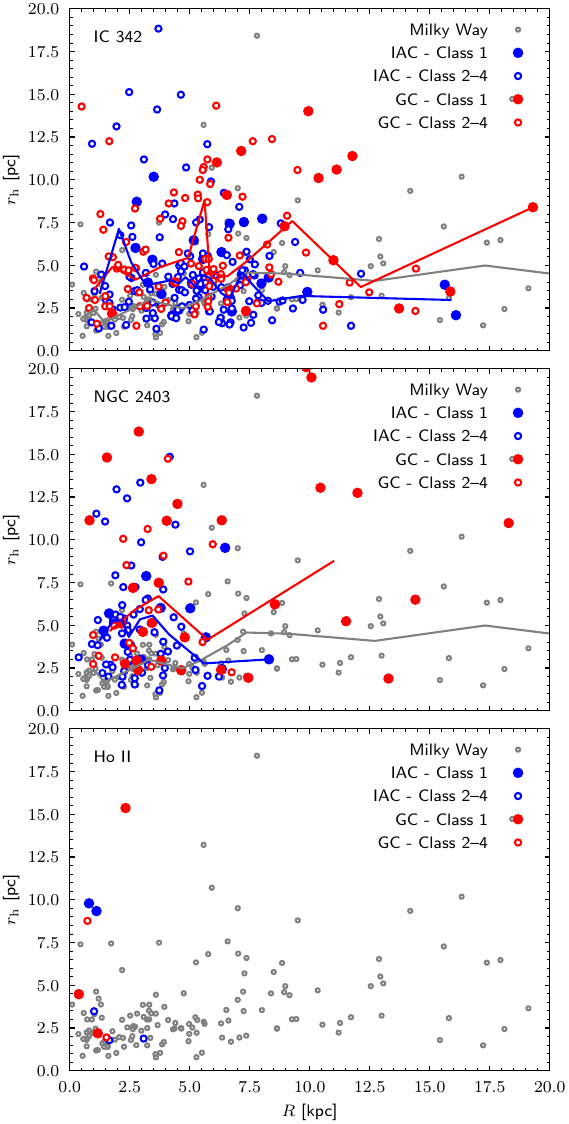}
\caption{Cluster half-light radii versus projected galactocentric distance for globular and young clusters, compared with Milky Way GCs. The coloured lines show median values for the corresponding types of clusters in bins of 10.}
\label{fig:sizes_gc_R}
\end{figure}

Figure~\ref{fig:sizes_gc} shows the half-light radii as a function of $M(\IE)$ for cluster candidates in each of the three galaxies.
Symbols are again colour-coded according to the classification of the clusters as GC (red) or IAC (blue) candidates. We have omitted the unclassified objects and the class 10 objects.  
For reference, we also show the corresponding relations for Milky Way GCs, using the half-light radii from \citet{Harris1996}.

The old GC candidates brighter than $M(\IE) \approx -8.5$ generally have half-light radii in the range 2--5\,pc, similar to what is seen for the Milky Way GCs. At fainter magnitudes, however, the spread in the sizes increases considerably, with a tail up to $r_\mathrm{h}\approx15$\,pc. By comparison with Fig.~\ref{fig:fwhm_hst_euclid},
it is clear that this scatter cannot be attributed to uncertainties on the size measurements. Nor is the scatter driven by sources with uncertain classifications;  
many of the largest candidates are high-confidence class 1 clusters.
While the Milky Way does host a number of similarly large GCs, they are rarer than in either of the two spirals studied here. 
A Kolmogorov--Smirnov test confirms that the size distributions of old GC candidates in both NGC\,2403 and IC\,342 differ significantly from that of the Milky Way GCs, with K--S $p$-values of $2\times10^{-6}$ and $10^{-7}$, respectively. The difference between NGC\,2403 and IC\,342 is somewhat less significant, at $p=0.04$.

Figure~\ref{fig:sizes_gc_R} shows the measured cluster sizes as a function of the projected galactocentric distances. The coloured symbols have their usual meaning and in addition the solid lines connect the median values in bins of 10 clusters. 
As is well known \citep{VanDenBergh1991,Baumgardt2018}, there is a correlation between size and Galactocentric distance for Milky Way GCs, and although this trend becomes somewhat weaker when projected onto the 2D plane, it is still noticeable in Fig.~\ref{fig:sizes_gc_R}. It is less evident that such a trend is present in our data, and the median sizes are seen to be larger than in the Milky Way, driven upwards by the larger fraction of extended clusters. 
It is, however, somewhat striking that no clusters larger than about 10\,pc are found in IC\,342 outside a projected galactocentric distance of $\sim$12\,kpc, coinciding roughly with the outer edge of the IC\,342 disc \citepalias{ERONearbyGals}. This may suggest that the fainter, extended clusters constitute a disc population that is distinct from the classical GCs.
In this sense, NGC\,2403 is different since it does have a number of relatively large clusters beyond the point where the surface brightness of the disc starts to drop rapidly, at about 5\,kpc \citepalias{ERONearbyGals}.

\section{Discussion}

\subsection{Globular clusters: specific frequencies and luminosity functions}
\label{sec:gcs}

\begin{figure}
\centering
\includegraphics[width=\hsize]{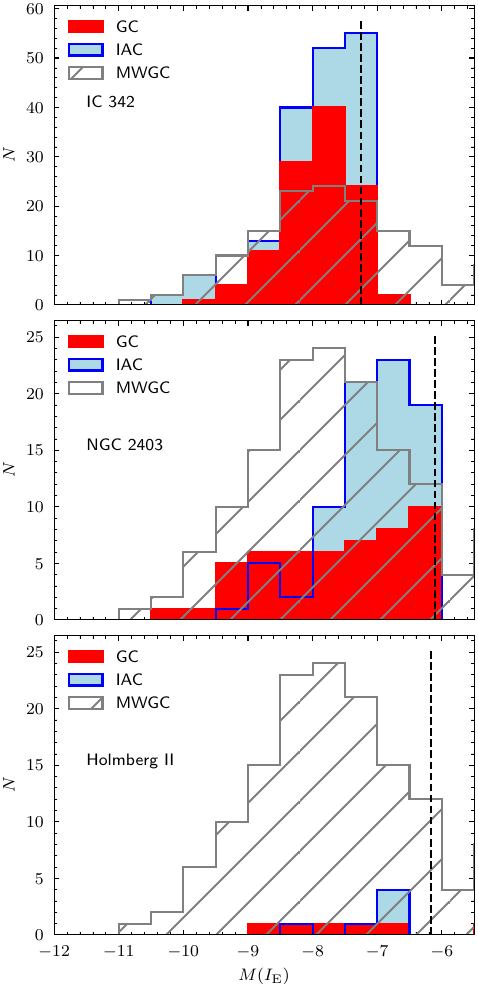}
\caption{Luminosity functions of class 1--4 cluster candidates in each of the three galaxies, compared with the Milky Way GCLF. The dashed lines indicate the $\IE=21.5$ selection limits for the cluster candidates.}
\label{fig:LFs_cfhtsel}
\end{figure}

When calculating GC specific frequencies, 
a common approach to circumventing the difficulty of identifying the faintest clusters is to assume that the GCLFs follow the roughly Gaussian distributions as a function of magnitude commonly observed, and then to double the number of objects brighter than the peak \citep{Harris1981}. In NGC\,2403, IC\,342, and Holmberg\,II there are 19, 45, and 2 old GC candidates with $M(\IE) < -8$, so that the estimated total populations would then be $\sim38$, $\sim90$, and $\sim4$ GCs in each galaxy.
For the $M_V$ values in Table~\ref{tab:overview},
the corresponding specific frequencies are then $S_N = 0.81$ (NGC\,2403), $S_N = 0.27$ (IC\,342), and $S_N=0.62$ (Holmberg\,II). These are fairly typical values, albeit somewhat on the low side for IC\,342.
This does not account for any clusters that might lie outside the \Euclid\ footprint, but this would probably be a minor correction for NGC\,2403 and Holmberg\,II, and even for IC\,342 if the radial structure of its GC system is similar to that of the Milky Way. However, we could be missing a more extended component of the GC system, such as that associated with M31. 

The fact that the total numbers of GCs actually detected in each galaxy (Table~\ref{tab:ccsel}) \emph{exceed} those estimated in the previous paragraph
suggests that the simple characterisation in terms of total GC numbers and $S_N$ may be somewhat deceptive. Indeed, comparing the GCLFs in more detail (Fig.~\ref{fig:LFs_cfhtsel}), it becomes clear that neither the NGC\,2403 nor IC\,342 GCLFs match that of the Milky Way particularly closely (for Holmberg\,II the small number of candidates makes such a comparison less conclusive).  
The requirement that candidates must pass our visual inspection may contribute to a decrease in completeness towards the selection limits (dashed lines in Fig.~\ref{fig:LFs_cfhtsel}), despite the fact that our tests in Sect.~\ref{sec:artclust} suggest a better than 90\% detection efficiency all the way down to $\IE=21.5$. With this in mind, we conclude that the LFs of neither IAC nor GC candidates show a clearly detected peak or turn-over, although such a detection would have been expected, at least for NGC\,2403, based on comparison with the Milky Way GCLF. Here it is important to recall that our GC candidate samples may include younger objects scattering across the boundaries defining the separation between IAC and GCC samples (Fig.~\ref{fig:TCDs_ug_gr}), particularly at the faint end where the photometric uncertainties increase. 

For IC\,342, however, the \emph{bright} end of the GCLF shows a real and significant deficit of GCs compared to the Milky Way.  The Milky Way has 30 GCs that would have $I_{\mathrm{E},0} \lesssim 19$, or $M(\IE) \lesssim -9.7$, at the distance of IC\,342, while we find only 11 old GC candidates in this magnitude range in IC\,342. 
A naive scaling based on these numbers would then suggest that IC\,342 hosts only $\sim11/30\times157 = 58$ ``true'' GCs, an even smaller number than that obtained by doubling the number of clusters brighter than the GCLF turn-over, and implying an even lower $S_N = 0.18$. It thus appears that the GCLF in IC\,342 may indeed be somewhat ``bottom-heavy'' compared to that of the Milky Way. This may be partly due to the same effect noted for NGC\,2403, that is, the sample of GC candidates being contaminated by younger objects and/or background galaxies.  We recall that other properties of the fainter candidates suggest that some fraction of them might not be classical GCs (Sect.~\ref{sec:sizes}).

As there is no sharp separation between the regions of the $(u-g, g-r)$ plane (Fig.~\ref{fig:TCDs_ug_gr}) defining our IAC and GCC sources, we briefly discuss the sensitivity of the preceding results to the colour criteria employed. In particular, the SSP models in Fig.~\ref{fig:SSPcol_age} suggest that a bluer colour limit, e.g. at $(u-g)_0=0.6$, would lead to a more complete sample of metal-poor GC candidates. At the same time, however, this would also be expected to increase the fraction of intermediate-age interlopers in our GC candidate sample. 
In fact, changing the $(u-g)_0$ limit to 0.6 mainly affects the statistics of the fainter part of the GCC samples: the total numbers of GC candidates increase from 111 to 130 (IC\,342) and from 50 to 52 (NGC\,2403), while the number of GC candidates brighter than $M(\IE)=-8$ increases slightly from 45 to 49 in IC\,342 and remains unchanged at 19 in NGC\,2403. In IC\,342, the number of GC candidates with $I_E<19$ remains unchanged at 11.  

Our study would not be the first to suggest an excess of relatively faint, GC-like objects; this has also been found by HST-based studies of other spirals, such as NGC\,6946, M101, NGC\,628, and NGC\,3627, where these fainter objects appear to be associated with the discs of their respective galaxies \citep{Chandar2004,Barmby2006,Simanton2015,Floyd2024}.
The diversity of cluster demographics is also illustrated by the ``faint fuzzy'' or ``diffuse'' star clusters that have been identified in a number of S0-type galaxies \citep{Brodie2002,Peng2006,Hwang2006}, as well as in M31 \citep{Huxor2005,Huxor2014}, NGC\,6822 (Howell et al., in prep.; \citealt{Hwang2011}) and other galaxies. These again tend to be fainter than the GCLF turn-over and have half-light radii in the range 7--15\,pc, not unlike the excess fainter clusters that we have identified in IC\,342 and, in smaller numbers, in NGC\,2403. 
The relatively large sizes of several of the GCs located outside the main disc of NGC\,2403 are also reminiscent of those measured for GCs in the outskirts of M33 \citep{Cockcroft2011}.
More reliable estimates of the ages and other properties of the clusters will require follow-up spectroscopy or deeper, multi-colour space-based imaging.

Capitalising on \Euclid's unique suitability for mapping the outer regions of GC systems, we can also compare the number of GC candidates in IC\,342 beyond the $R=10$\,kpc circle indicated in Fig.~\ref{fig:ic342cc_yc_gc} with statistics for the Milky Way and M31. Outside the 10\,kpc circle, we find 13 GC candidates in IC\,342. Using the data from Sect.~\ref{sec:compl} to project the simulated Milky Way GC system onto the same part of the \Euclid\ footprint, we find a very similar number of GCs (12) brighter than our $\IE=21.5$ selection limit. For the brightest clusters $(\IE<19)$, for which the IC\,342 GC system as a whole has a deficit with respect to that of the Milky Way, the comparison for the outer regions is less conclusive: there are two such GCs in the Milky Way and none in IC\,342. For comparison, M31 has 78 GCs with projected radii between 10\,kpc and 20\,kpc and estimated \IE\ magnitudes brighter than 21.5 (here assuming $V-\IE=0.5$) if shifted to the distance and extinction of IC\,342. Hence, while the data allow for the existence of a GC system in the outer regions of IC\,342 roughly comparable to what is found in the Milky Way, a GC system as rich as that associated with M31 appears to be clearly ruled out.
NGC\,2403, finally, has a total of seven GC candidates outside 10\,kpc, a fairly similar number to the five found in M33 \citep{Cockcroft2011}. 

\subsection{Globular clusters: colours and metallicities}
\label{sec:gcc_cm}

\begin{figure}
\centering
\includegraphics[width=\hsize]{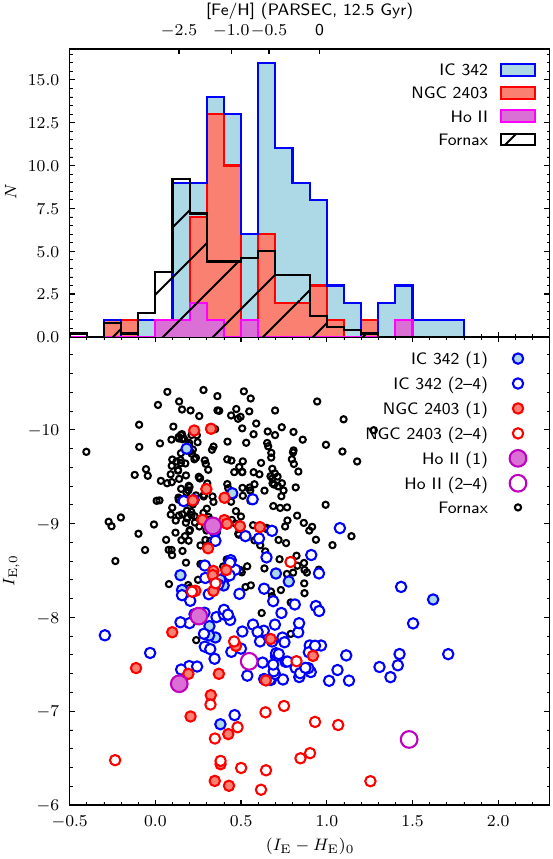}
\caption{Colour distributions of GC candidates in the three Showcase galaxies. Also included are data for spectroscopically confirmed GCs in the Fornax cluster \citep{EROFornaxGCs}.}
\label{fig:GCcol}
\end{figure}

Figure~\ref{fig:GCcol} shows the \Euclid\ CMDs and colour distributions for the old GC candidates in each galaxy together with
 \Euclid\ photometry for spectroscopically confirmed GCs in the Fornax cluster \citep{EROFornaxGCs}.
On the top axis we indicate the colours corresponding to  $\mathrm{[Fe/H]}=-2.5, -1.0, -0.5$, and $0$, according to the PARSEC SSP models. Here we have used models for an age of 12.5\,Gyr, recalling that the \Euclid\ colours depend only weakly on age (Fig.~\ref{fig:SSPcol_age}). 

The colour distribution of the Fornax GCs extends to somewhat bluer colours than in the three Showcase galaxies.  
This may be partly due to the colour cuts applied when selecting the Showcase GC candidates: these cuts were intended to reduce contamination by younger clusters, but may also have eliminated some genuine old, very metal-poor GCs. However, changing the selection limit to $(u-g)_0=0.6$ does not significantly modify the $(\IE-\HE)_0$ colour distributions of our GC candidates.
There are about 10 sources with much redder colours, $(\IE-\HE)_0>1.2$, than predicted by the SSP models for any age/metallicity combination, most of which are associated with IC\,342. 
We have inspected Herschel SPIRE and PACS images of IC\,342 and found no particularly prominent features that might suggest a strong increase in Galactic foreground extinction at the locations of these objects. 
We also attempted to apply corrections for local variations in the reddening, based on the colours of RGB stars (Annibali et al., in prep.) but found this to have a negligible impact on our results, much less being able to account for the very red sources in IC\,342. 
However, most of them tend to be located in dense regions of the IC\,342 disc and it is possible that some are heavily reddened, younger objects. For example, one of these sources, ESCC-IC342-091, a high-confidence cluster candidate (Figs.~\ref{fig:ic342stamps1gc} and \ref{fig:hst_ic342stamps1gc}), is associated with enhanced emission in a WISE 12 $\mu$m image, where PanSTARRS colour images also suggest the presence of significant dust extinction. Other sources may be misclassified background galaxies. We also note, from the artificial cluster experiments (Fig.~\ref{fig:i342_mwgc_cmd}) that a small number of clusters scatter far from the input colours, hence effects of photometric uncertainty cannot be entirely discounted.

Within the range of ``normal'' GC colours, some differences between the four systems may be noted. 
The colour distribution for GC candidates in IC\,342 appears to be more heavily weighted towards relatively red colours. Comparing with the CMD in the lower panel, we see that this is mainly due to the fainter objects discussed previously, and again suggests that these might be relatively low-mass, metal-rich and/or reddened objects associated with the disc. In all three galaxies, the histograms suggest a peak in the colour distribution around $(\IE-\HE)_0\approx0.3$, while IC\,342 and NGC\,2403 also have peaks at redder colours. A KMM test for bimodality \citep{Ashman1994} confirms the visual impression that the colour distributions for GC candidates in both spirals are better approximated by a sum of two Gaussians than by a single Gaussian. For candidates in the range $0<(\IE-\HE)_0<1.2$, the KMM test returns $p$-values of 0.003 and 0.001, respectively, with peaks at $(\IE-\HE)_0 = 0.36$ and 0.79 (IC\,342) and 0.36 and 0.80 (NGC\,2403). Converting these colours to metallicities using the 12.5\,Gyr PARSEC models, we find $\mathrm{[Fe/H]}=-1.26$ and $\mathrm{[Fe/H]}=-0.29$ for IC\,342 and 
$\mathrm{[Fe/H]}=-1.25$ and $\mathrm{[Fe/H]}=-0.27$ for NGC\,2403. A correction for an assumed $\alpha$-enhanced composition of the GCs would decrease these values by about 0.2\,dex, and the mean metallicities of the two populations would then be fairly similar to the mean metallicities of the metal-poor and metal-rich GCs in the Milky Way \citep{Zinn1985} and other large galaxies \citep{Larsen2001b,Peng2006}.
The effect of changing the colour limit for GC candidate selection to $(u-g)_0=0.6$ is minimal: the mean colours returned by the KMM analysis change by only 0.01\,mag.

In Holmberg\,II, seven GC candidates have LBT colours suggesting they might be old GCs. Three of these are included in the LEGUS catalogue with poorly constrained ages (our IDs ESCC-HoII-015, ESCC-HoII-013, and ESCC-HoII-012, with LEGUS IDs \#4220, \#3, and \#11, respectively).
The faint, outlying objects ESCC-HoII-001 and ESCC-HoII-002, both classified by us as class~1 GC candidates, do not appear to have been noticed previously. Both have a $(g-r)_0 = +0.44$ with uncertainties of 0.01--0.02\,mag, while the \Euclid\ colours are $(\IE-\HE)_0 = 0.257 \pm 0.014$ and $0.028 \pm 0.019$, respectively, consistent with them being old, metal-poor GCs.
Both are quite extended with $r_\mathrm{h}=15$\,pc and $30$\,pc, respectively, raising the question whether they are more properly classified as stellar clusters or perhaps rather as ultra-faint dwarf galaxies (UFDs). With absolute magnitudes of $M(\IE) = -5.3$ and $-4.8$, they are about 3\,mag fainter the peak of the GCLF, and thus located in the region of the size-luminosity space where properties of GCs and UFDs overlap \citep{Drlica-Wagner2015,Simon2019}.
Of the two Local Group ERO targets, NGC\,6822 has about five extended clusters with $r_\mathrm{h}>10$\,pc (including one object newly discovered in the ERO data), of which two are faint objects with magnitudes comparable to those of the two objects in Holmberg\,II \citep[][Howell et al., in prep.]{Hwang2011,Huxor2013}. No such clusters have so far been found in IC\,10. 

\subsection{The young star cluster systems}

\begin{figure}
\centering
\includegraphics[width=\hsize]{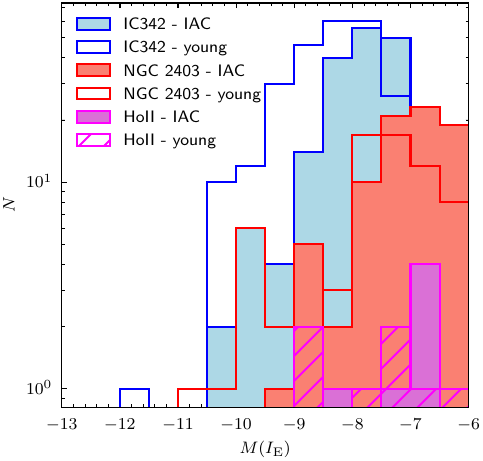}
\caption{Luminosity functions of young star cluster candidates.}
\label{fig:LFcmp}
\end{figure}

Finally turning briefly to the younger cluster candidates, we show the luminosity functions (LFs) for the IAC and young candidates in Fig.~\ref{fig:LFcmp}. The filled and open histograms show the LFs for the class 1--4 IAC candidates and the class~10 candidates, respectively. 
From a maximum-likelihood fit of a power-law to the LFs of the IAC candidates in IC~342 with $M(\IE) < -8$, $\mathrm{d}N/\mathrm{d}L \propto L^\alpha$, we find a slope of $\alpha=-2.64\pm0.26$, and $\alpha=-2.37\pm0.10$ for the combined young/intermediate-age sample.
Combining the IAC and young clusters brighter than $M(\IE)=-8$ in all three galaxies, we get $\alpha=-2.32\pm0.09$.
These slopes are similar to those reported for young cluster populations in a variety of other star-forming galaxies \citep[e.g.][]{Larsen2002,Whitmore2014}.  Such relatively steep cluster LFs are consistent with a scenario in which the underlying cluster mass function has the shape of a power-law with an exponent close to $-2$ at low masses, with an exponential truncation at a mass that depends on the intensity of star formation, but is typically in the range of several times $10^4 M_\odot$ to a few times $10^5 M_\odot$ in normal star-forming galaxies \citep{Larsen2009,PortegiesZwart2010,Johnson2017}.

While no particular effort has been made to ensure or quantify the completeness of the young cluster samples, 
it is interesting to compare the statistics for candidates in Holmberg\,II and NGC\,2403 with their Local Group analogues, the SMC and M33. 
The PHATTER cluster catalogue for M33 includes 22 clusters with $M_\mathrm{F814W} < -8$ \citep{Johnson2022}, where we have converted the Vega magnitudes in the PHATTER catalogue to AB magnitudes for consistency with the \Euclid\ \IE\ magnitudes. This is just slightly less than our combined sample of $8+18=26$ intermediate-age and young candidates in NGC\,2403, consistent with the two galaxies having roughly similar star-formation rates. 
For the SMC, we use the cluster catalogue by \citet{Rafelski2005} and convert their $I$ magnitudes to the AB system by adding 0.45\,mag \citep{Blanton2007}. In this way we find 9 SMC clusters with $M_{I,AB} < -8$. The difference with respect to the three young+IAC clusters in Holmberg\,II may be partly explained by the somewhat higher star-formation rate of the SMC (Sect.~\ref{sec:overview}). 
In addition to homogenising the photometric data, a more refined analysis should also adopt a uniform strategy for identification of young clusters in different galaxies. 

\section{Summary and conclusions}

We have presented an analysis of the star cluster populations in three nearby galaxies, IC\,342, NGC\,2403, and Holmberg\,II, observed as part of the \Euclid\ Early Release Observations programme. For IC\,342, this is the first study of stellar clusters (other than the nuclear cluster).
Our main focus is on the old clusters, where the \Euclid\ observations have the greatest potential for improvement over existing studies as they allow a more complete census of the cluster populations also in the outer regions of galaxies. 
At distances of 3.20--3.45\,Mpc, the \Euclid\ field-of-view covers a region of $39\,\mathrm{kpc}\times45\,\mathrm{kpc}$ to $42\,\mathrm{kpc}\times48\,\mathrm{kpc}$ around each galaxy, which is large enough to include about 90\% of the known Milky Way GC system if placed at the distance of our target galaxies (although we note that the M31 GC system has about 100 GCs, or about 20\% of the entire population, falling outside 20\,kpc).  

Our main results are as follows:
\begin{itemize}
\item We identify 111, 50, and 7 old GC candidates in IC\,342, NGC\,2403, and Holmberg\,II, respectively. 
These numbers include 15 previously identified GC candidates in NGC\,2403 (including 8 candidates from \citetalias{ERONearbyGals}) and two faint, extended ($r_\mathrm{h}=15$\,pc and 30\,pc), and previously unknown stellar systems outside the main body of Holmberg\,II. With $M(\IE)=-5.3$ and $-4.8$, it is unclear whether the latter are stellar clusters or UFDs. 
\item The median (projected) galactocentric distances of the GC candidates in IC\,342 and NGC\,2403 are 5.7\,kpc and 3.7\,kpc, respectively. Compared with the corresponding values for the Milky Way (5.0\,kpc) and M31 (7.0\,kpc), these values thus appear fairly typical. 
\item 
By doubling the numbers of GCs brighter than $M(\IE)=-8$, corresponding to the expected peak of the GC luminosity function, we find specific frequencies of $S_N = 0.28$ (IC\,342), 0.81 (NGC\,2403), and 0.62 (Holmberg\,II). The values for NGC\,2403 and Holmberg\,II are fairly typical for late-type galaxies, but somewhat on the low side for IC\,342. The total GC population of NGC\,2403 also agrees well with expectations based on the relation between galaxy mass and GC numbers \citep{Burkert2020,Forbes2022}.
\item
In both IC\,342 and NGC\,2403, we find an excess of relatively faint clusters compared to the canonical, approximately Gaussian, shape of the GC luminosity function, as reported previously for other spiral galaxies. The faint, extended, predominantly red clusters in IC\,342 are largely confined to the disc and resemble the ``faint fuzzy'' star clusters observed in the discs of some lenticular galaxies. NGC\,2403, by contrast, also hosts a number of relatively extended clusters beyond the disc, as found previously for M31 \citep{Huxor2014} and M33 \citep{Cockcroft2011}.
\item The \Euclid\ $\IE-\HE$ colours are consistent with the GC candidates spanning a broad range of metallicities, $-2.5 \lesssim \mathrm{[Fe/H]} \lesssim 0$. Both IC\,342 and NGC\,2403 exhibit bimodal colour distributions with mean metallicities similar to those of the metal-poor and metal-rich GC subpopulations observed in the Milky Way and other large galaxies. 
\item From a power-law fit to the combined LFs of the young and intermediate-age candidates, we find a slope of $-2.32\pm0.09$, similar to what has been reported in previous studies of young cluster populations in a variety of galaxies. 
\end{itemize}

Despite the relatively small sample of galaxies investigated here, the comparison with their Local Group analogues have yielded intriguing hints that interesting trends might emerge from studies of larger samples. On the one hand, the GC systems of M33 and NGC\,2403 appear similar in many respects, including the total number of GCs, their spatial distributions, and the presence of a number of relatively extended clusters in the outskirts of the galaxies. In contrast, the GC system of IC\,342 differs more significantly from those of the Milky Way and M31, in particular regarding the relatively small number of luminous GCs present in IC\,342.
If we adopt the view that a significant fraction of the GC populations around major galaxies are associated with accretion events, we might speculate that IC\,342 has had an unusually quiescent accretion history. It would be of great interest to investigate the characteristics of the stellar halo, for example in terms of its spatial- and metallicity distributions, and compare with those of its Local Group counterparts.

The preceding analysis demonstrates the power of the large field-of-view, combined with exquisite image quality offered by \Euclid, for studies of nearby galaxies.
The full Euclid Wide Survey will not be limited by the coverage of a single observation but will provide contiguous coverage of thousands of square degrees of sky, including many galaxies in the Local Universe where studies similar to that presented here can be undertaken. 
The analysis presented here also demonstrates the powerful combination of the high resolution imaging provided by \Euclid\ with ground-based multi-colour photometry, not only for the core science but also for legacy science on stellar populations. The results presented in this paper should then not be seen as definitive, but rather as a preview of things to come.

\begin{acknowledgements}
  
\AckEC  

AMNF acknowledges STFC grant ST/Y001281/1
This work has made use of the Early Release Observation (ERO) data from the Euclid mission of the European Space Agency (ESA), 2024, https://doi.org/10.57780/esa-qmocze3

\end{acknowledgements}

%
%

\bibliography{bibtex,Euclid}

%

\begin{appendix}
  \onecolumn 
  
\section{Cluster candidates}

\begin{table}[h!]
\caption{\Euclid\ star cluster candidates.  
}
\setlength{\tabcolsep}{3.25pt}
\smallskip
\label{table:i342cc}
\smallskip
\small
\begin{tabular}{ccccccccccc} \hline
  & \\[-7pt]
 ID & Master ID & Other ID & Class & RA (2000.0) & Dec (2000.0) & FWHM & $B/A$ & \\
 & \IE & \YE & \JE & \HE & $u$ & $g$ & $r$ \\
 & & & & & & & & & \\[-8pt]
\hline
& & & & & & & & & \\[-8pt]
ESCC-NGC2403-001 & 374860 &  ESCC1 &  1 (GCC) & 113.6659832 & +65.8406444 &  4.78 & 0.85 \\
 & $21.17\pm 0.01$ & $20.75\pm 0.01$ & $20.71\pm 0.01$ & $20.73\pm 0.00$ & $22.87\pm 0.05$ & $21.77\pm 0.02$ & $21.27\pm 0.02$ \\
ESCC-NGC2403-002 & 259038 &  ESCC2 &  1 (GCC) & 113.6923724 & +65.6259126 &  5.55 & 0.86 \\
 & $19.32\pm 0.00$ & $18.93\pm 0.00$ & $18.90\pm 0.00$ & $18.92\pm 0.00$ & $21.13\pm 0.01$ & $20.09\pm 0.01$ & $19.50\pm 0.01$ \\
ESCC-NGC2403-003 & 360879 &  ESCC3 &  1 (GCC) & 113.8561006 & +65.7187595 &  5.68 & 0.87 \\
 & $20.66\pm 0.01$ & $20.31\pm 0.01$ & $20.32\pm 0.01$ & $20.39\pm 0.00$ & $22.38\pm 0.03$ & $21.29\pm 0.01$ & $20.82\pm 0.01$ \\
ESCC-NGC2403-004 & 313462 &  ESCC4 &  1 (NC) & 113.8722446 & +65.6541417 &  1.31 & 0.85 \\
 & $21.08\pm 0.01$ & $21.08\pm 0.01$ & $21.28\pm 0.02$ & $21.47\pm 0.00$ & $21.83\pm 0.03$ & $21.08\pm 0.01$ & $21.10\pm 0.02$ \\
ESCC-NGC2403-005 &  15230 &  ESCC5 &  1 (GCC) & 114.0401552 & +65.3527644 &  2.83 & 0.92 \\
 & $19.76\pm 0.00$ & $19.47\pm 0.01$ & $19.53\pm 0.01$ & $19.60\pm 0.00$ & $21.64\pm 0.02$ & $20.26\pm 0.01$ & $19.89\pm 0.01$ \\
ESCC-NGC2403-006 &  51661 &  ESCC6 &  3 (GCC) & 114.4157049 & +65.5158114 &  0.98 & 0.96 \\
 & $19.86\pm 0.00$ & $19.44\pm 0.01$ & $19.38\pm 0.01$ & $19.34\pm 0.00$ & $21.80\pm 0.02$ & $20.60\pm 0.01$ & $20.08\pm 0.01$ \\
ESCC-NGC2403-007 &  29907 &  ESCC7 &  1 (GCC) & 114.5104579 & +65.4709785 &  8.49 & 0.84 \\
 & $20.20\pm 0.00$ & $19.88\pm 0.01$ & $19.89\pm 0.01$ & $19.95\pm 0.00$ & $21.93\pm 0.02$ & $20.82\pm 0.01$ & $20.30\pm 0.01$ \\
ESCC-NGC2403-008 &  58158 &  ESCC8 &  1 (GCC) & 114.5153481 & +65.5153979 &  2.71 & 0.88 \\
 & $19.32\pm 0.00$ & $18.98\pm 0.01$ & $18.97\pm 0.01$ & $19.03\pm 0.00$ & $20.86\pm 0.01$ & $19.87\pm 0.01$ & $19.45\pm 0.01$ \\
ESCC-NGC2403-009 &  46808 &  ESCC9 &  1 (GCC) & 114.5744057 & +65.5099033 & -99.00 & -99.00 \\
 & $21.34\pm 0.01$ & $21.01\pm 0.01$ & $20.92\pm 0.01$ & $20.94\pm 0.00$ & $23.23\pm 0.07$ & $22.04\pm 0.03$ & $21.52\pm 0.03$ \\
ESCC-NGC2403-010 & 366220 &     D6 &  1 (GCC) & 113.7740279 & +65.7575610 &  0.82 & 0.96 \\
 & $18.60\pm 0.00$ & $18.15\pm 0.00$ & $18.11\pm 0.00$ & $18.13\pm 0.00$ & $20.45\pm 0.01$ & $19.40\pm 0.00$ & $18.70\pm 0.00$ \\
ESCC-NGC2403-011 & 361920 &     F1 &  1 (GCC) & 113.8019055 & +65.7209848 &  2.28 & 0.75 \\
 & $18.56\pm 0.00$ & $18.19\pm 0.00$ & $18.18\pm 0.00$ & $18.23\pm 0.00$ & $20.32\pm 0.01$ & $19.32\pm 0.00$ & $18.74\pm 0.00$ \\
ESCC-NGC2403-012 & 184817 &     C3 &  1 (IAC) & 113.9289148 & +65.5921973 &  4.15 & 0.78 \\
 & $18.99\pm 0.00$ & $19.02\pm 0.01$ & $19.19\pm 0.01$ & $19.37\pm 0.00$ & $19.43\pm 0.00$ & $18.92\pm 0.00$ & $18.92\pm 0.00$ \\
ESCC-NGC2403-013 &  90028 &        & 10 (YC) & 113.9390452 & +65.5478183 &  1.48 & 0.57 \\
 & $20.75\pm 0.01$ & $21.02\pm 0.01$ & $21.15\pm 0.01$ & $21.13\pm 0.00$ & $20.59\pm 0.01$ & $20.70\pm 0.01$ & $20.63\pm 0.01$ \\
ESCC-NGC2403-014 & 268458 &        & 10 (YC) & 113.9591640 & +65.6299970 &  1.39 & 0.35 \\
 & $21.21\pm 0.01$ & $21.67\pm 0.03$ & $21.97\pm 0.04$ & $21.94\pm 0.00$ & $20.38\pm 0.02$ & $20.54\pm 0.03$ & $20.54\pm 0.03$ \\
ESCC-NGC2403-015 & 264875 &        & 10 (YC) & 113.9593628 & +65.6292573 &  3.29 & 0.66 \\
 & $20.56\pm 0.01$ & $20.69\pm 0.01$ & $20.90\pm 0.02$ & $20.65\pm 0.00$ & $20.53\pm 0.02$ & $20.49\pm 0.02$ & $20.20\pm 0.02$ \\
ESCC-NGC2403-016 & 265847 &        & 10 (YC) & 113.9605876 & +65.6300640 &  0.52 & 0.57 \\
 & $20.59\pm 0.01$ & $20.90\pm 0.02$ & $21.14\pm 0.02$ & $21.25\pm 0.00$ & $19.83\pm 0.01$ & $19.94\pm 0.01$ & $20.10\pm 0.02$ \\
ESCC-NGC2403-017 & 324497 &        &  1 (IAC) & 113.9736692 & +65.6599677 &  1.06 & 0.90 \\
 & $19.61\pm 0.00$ & $19.55\pm 0.01$ & $19.64\pm 0.01$ & $19.69\pm 0.00$ & $20.03\pm 0.01$ & $19.68\pm 0.01$ & $19.66\pm 0.01$ \\
ESCC-NGC2403-018 & 270093 &        & 10 (YC) & 113.9780801 & +65.6314992 &  1.36 & 0.62 \\
 & $20.16\pm 0.01$ & $20.60\pm 0.02$ & $21.02\pm 0.02$ & $21.04\pm 0.00$ & $19.59\pm 0.01$ & $19.82\pm 0.01$ & $19.73\pm 0.01$ \\
ESCC-NGC2403-019 & 255111 &        &  3 (NC) & 114.0000863 & +65.6262356 &  4.06 & 0.94 \\
 & $21.28\pm 0.01$ & $21.45\pm 0.03$ & $21.62\pm 0.04$ & $21.82\pm 0.00$ & $22.07\pm 0.04$ & $21.36\pm 0.03$ & $21.29\pm 0.04$ \\
ESCC-NGC2403-020 & 318587 &        &  1 (IAC) & 114.0002545 & +65.6549993 &  1.87 & 0.95 \\
 & $20.19\pm 0.00$ & $19.94\pm 0.01$ & $19.97\pm 0.01$ & $19.95\pm 0.00$ & $21.20\pm 0.02$ & $20.57\pm 0.01$ & $20.27\pm 0.01$ \\
\hline
\end{tabular}
\tablefoot{
The first 20 entries of the catalogue are listed here for guidance regarding the contents. The full catalogue is available on-line. For NGC\,2403, the first nine entries repeat the numbering scheme of \citetalias{ERONearbyGals}, while the remaining entries are sorted in order of increasing RA.
The FWHM values are given in VIS pixels ($0\farcs1$), from King model fits corrected for the PSF. A value of $-99.00$ indicates that no FWHM could be measured. The $B/A$ values are the minor/major axis ratios of the model fits. 
}
\end{table}

\section{Locations of cluster candidates}

\begin{figure*}
\centering
\includegraphics[width=\textwidth]{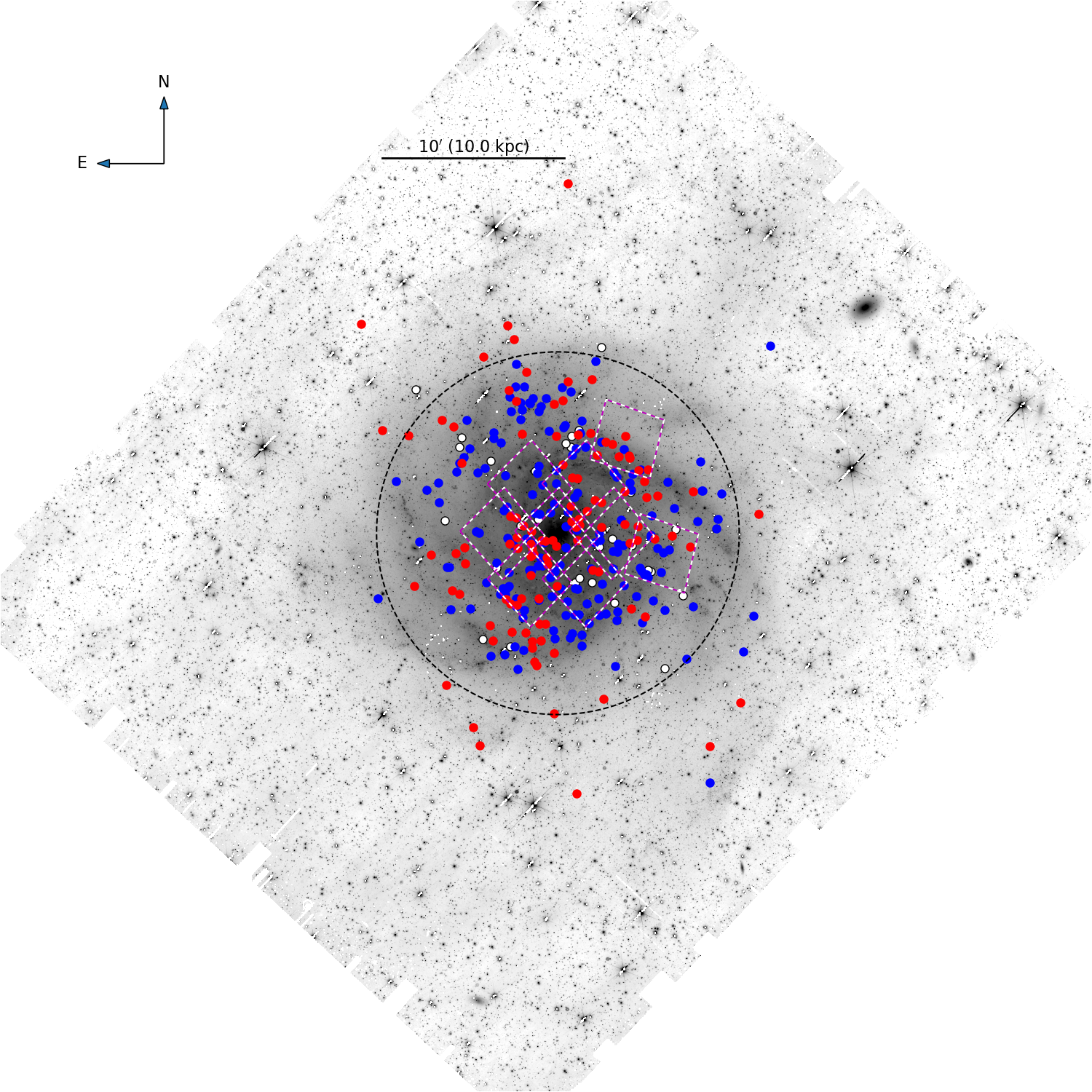}
\caption{\Euclid\ VIS image of IC\,342 with class 1--4 cluster candidates marked.  Old GC candidates are shown with red circles, intermediate-age candidates  in blue, and ambiguous candidates as open black circles. The dashed black circle indicates a radius of 10\,kpc and the white/magenta dashed lines indicate the HST/ACS F435W+F606W coverage from programmes 10579 and 16002. North is up and east to the left. }
\label{fig:ic342cc_yc_gc}
\end{figure*}

\begin{figure*}
\centering
\includegraphics[width=\textwidth]{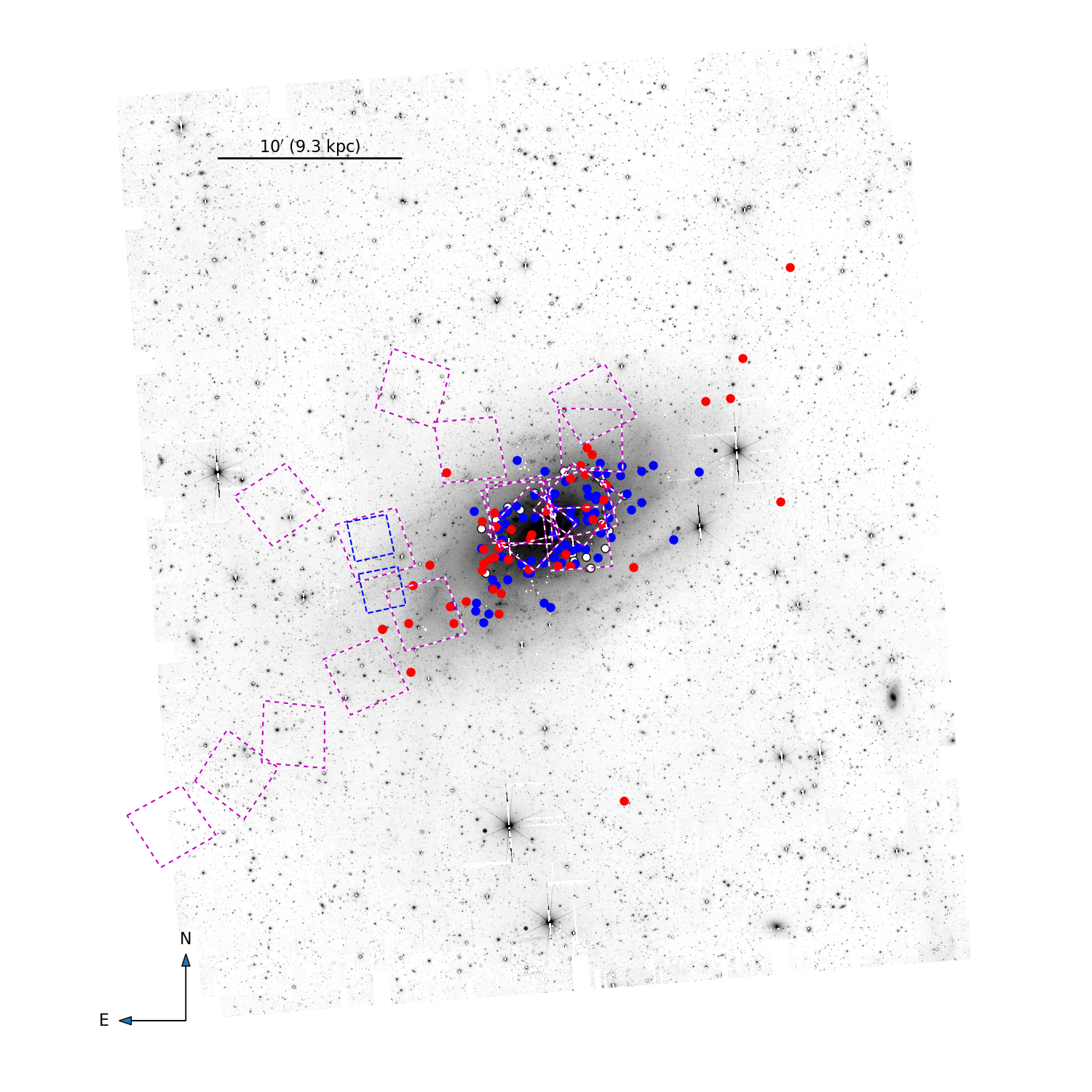}
\caption{\Euclid\ VIS image of NGC\,2403 with cluster candidates marked. Symbols and orientation of the image are as in Fig.~\ref{fig:ic342cc_yc_gc}. Also indicated is a JWST/NIRCAM pointing from Programme ID 1638 (dashed blue lines).}
\label{fig:ngc2403cc_yc_gc}
\end{figure*}

\begin{figure*}
\centering
\includegraphics[width=15cm]{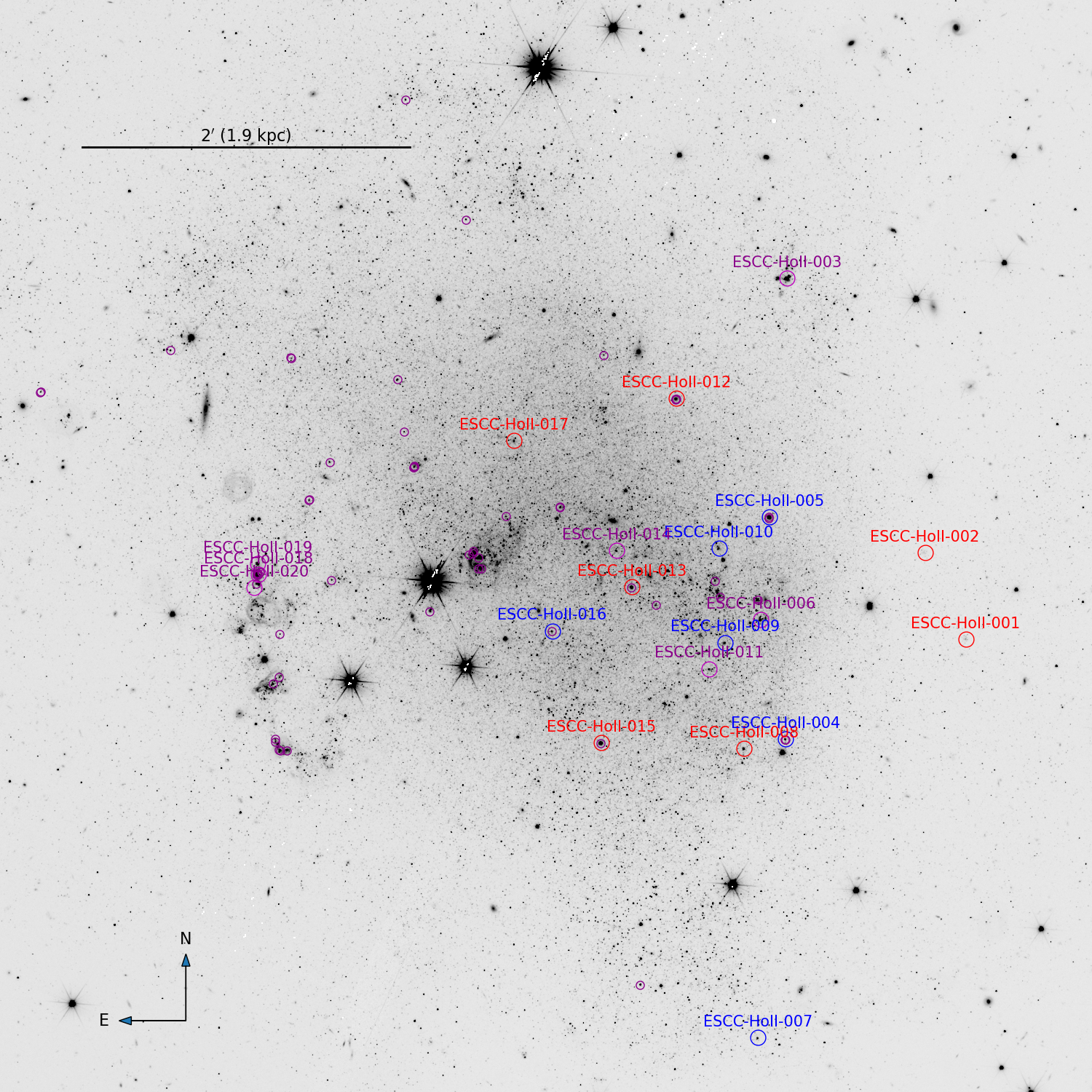}
\caption{\Euclid\ VIS image of the central $400\arcsec\times400\arcsec$ around Holmberg\,II with cluster candidates marked. 
As in Fig.~\ref{fig:ic342cc_yc_gc}, red and blue circles indicate our old GC and IAC candidates, respectively. Large magenta circles indicate young (class 10) objects and smaller magenta markers indicate clusters from the LEGUS survey \citep{Cook2019}.}
\label{fig:HoIIcc_yc_gc}
\end{figure*}

\section{Cutouts of cluster candidates}
\begin{figure}
\centering
\includegraphics[width=\hsize]{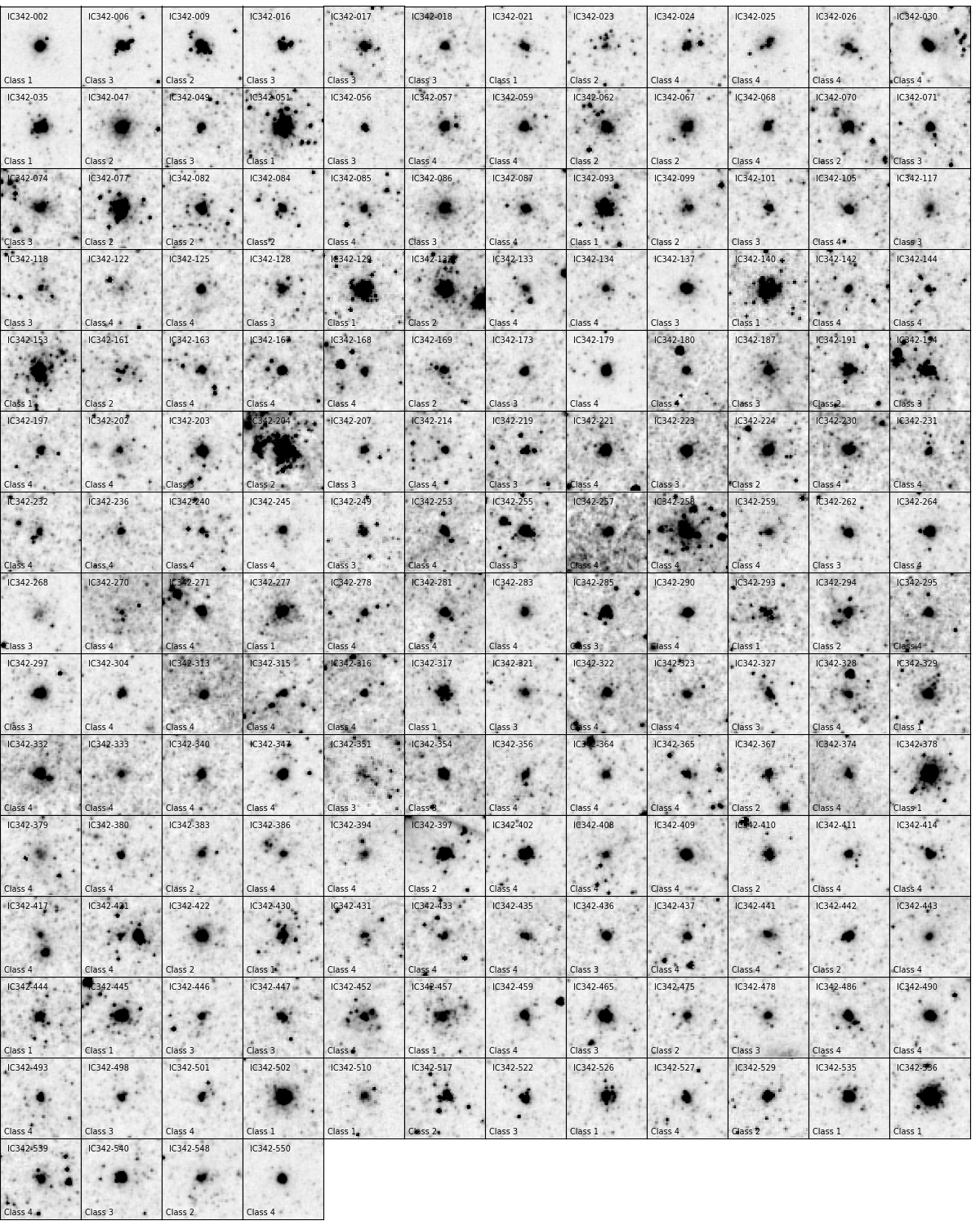}
\caption{Cut-outs of clusters in IC\,342 classified as intermediate-age candidates based on their MegaCam colours. The ID of each cluster is indicated with the leading ``ESCC'' stripped. Each cut-out measures $5\arcsec\times5\arcsec$.} 
\label{fig:ic342stamps1yc}
\end{figure}

\begin{figure}
\centering
\includegraphics[width=\hsize]{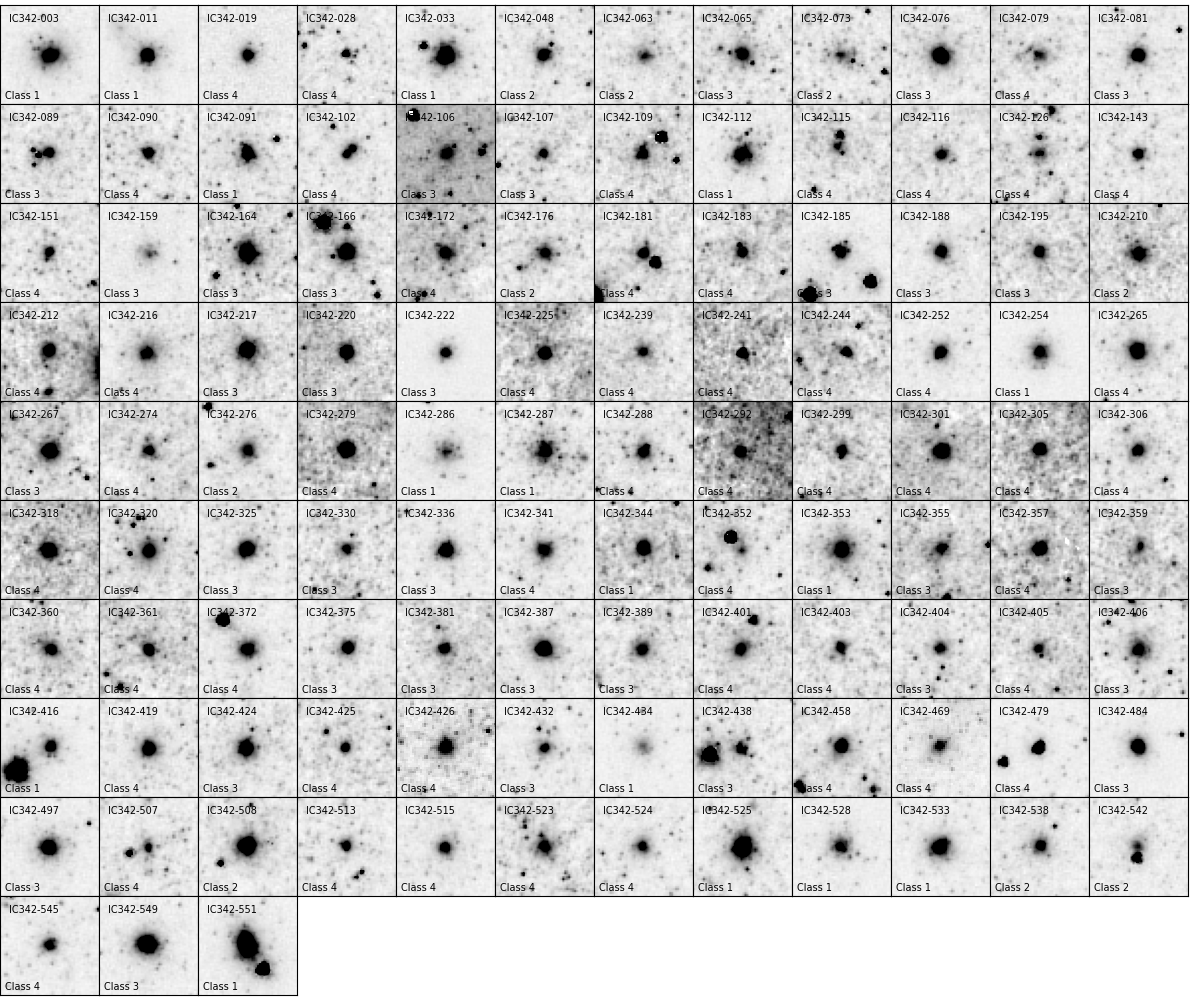}
\caption{Cut-outs of clusters in IC\,342 classified as GC candidates based on their MegaCam colours. Each cut-out measures $5\arcsec\times5\arcsec$.} 
\label{fig:ic342stamps1gc}
\end{figure}

\begin{figure}
\centering
\includegraphics[width=\hsize]{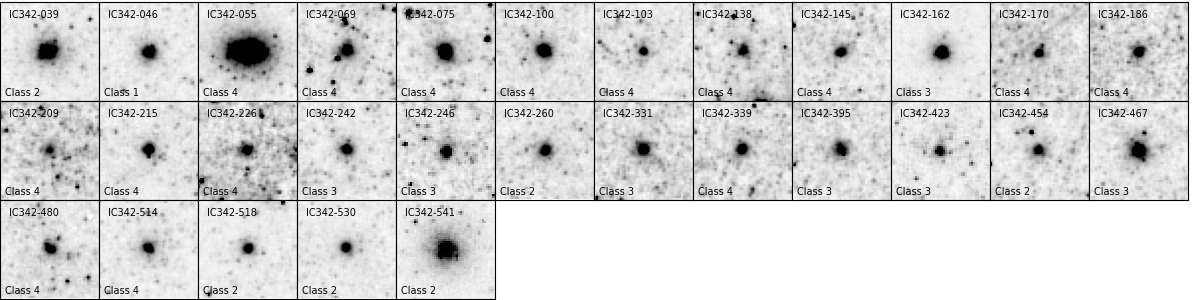}
\caption{Cut-outs of clusters in IC\,342 with ambiguous MegaCam colours. Each cut-out measures $5\arcsec\times5\arcsec$.} 
\label{fig:ic342stamps1nc}
\end{figure}

\begin{figure}
\centering
\includegraphics[width=\hsize]{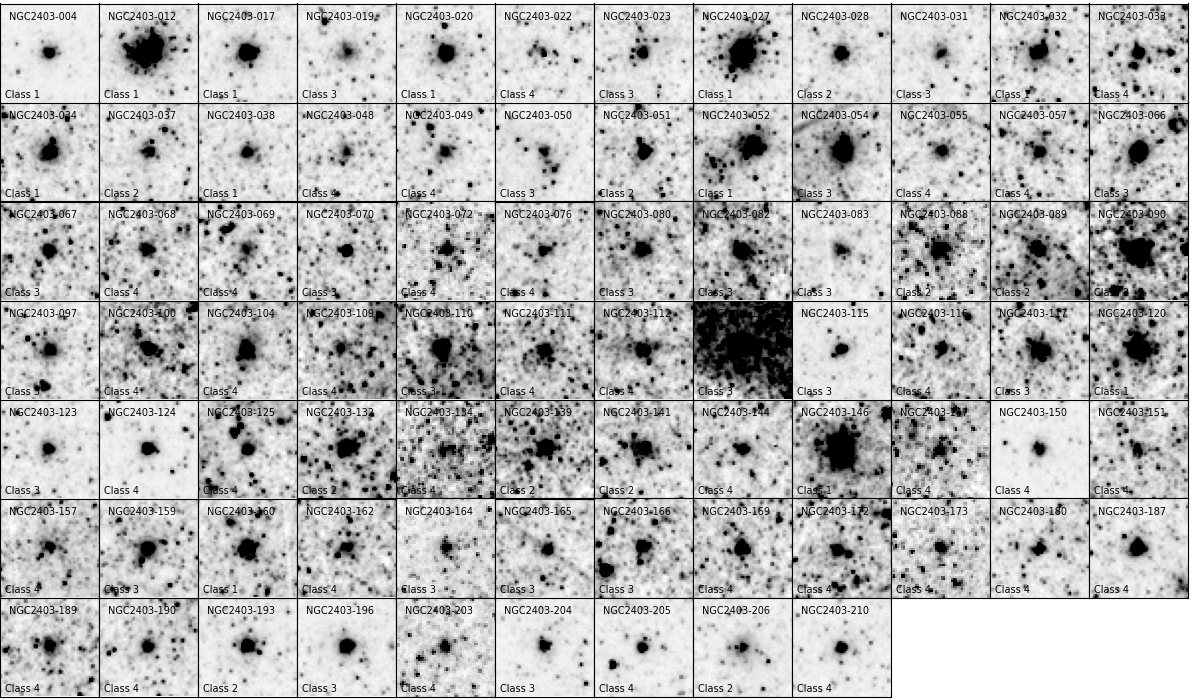}
\caption{Cut-outs of intermediate-age cluster candidates in NGC\,2403. Each cut-out measures $5\arcsec\times5\arcsec$.} 
\label{fig:ngc2403stamps1yc}
\end{figure}

\begin{figure}
\centering
\includegraphics[width=\hsize]{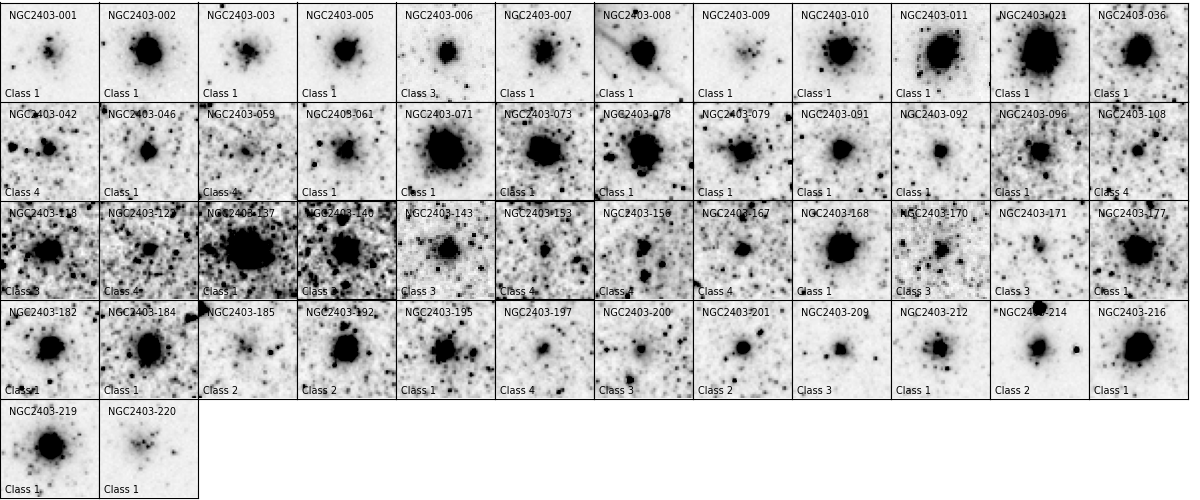}
\caption{Cut-outs of GC candidates in NGC\,2403. Each cut-out measures $5\arcsec\times5\arcsec$.} 
\label{fig:ngc2403stamps1gc}
\end{figure}

\begin{figure}
\centering
\includegraphics[width=\hsize]{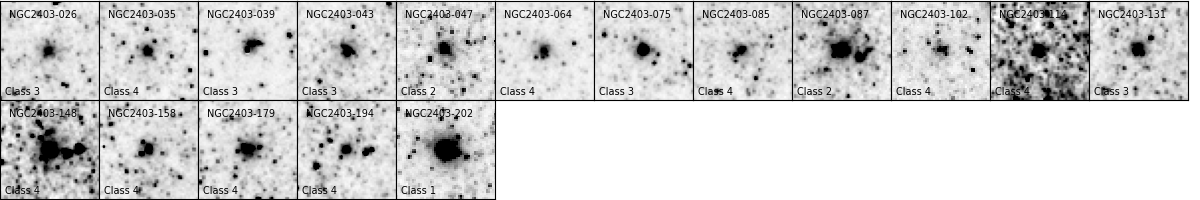}
\caption{Cut-outs of ambiguous cluster candidates in NGC\,2403.Each cut-out measures $5\arcsec\times5\arcsec$.} 
\label{fig:ngc2403stamps1nc}
\end{figure}

\begin{figure}
\centering
\includegraphics[width=140mm]{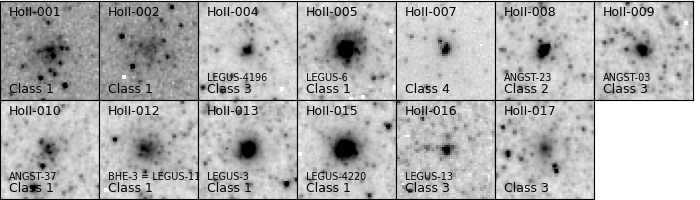}
\caption{Cut-outs of cluster candidates in Holmberg\,II. The objects HoII-001 and HoII-002 are shown at a higher contrast than the other objects. Each cut-out measures $5\arcsec\times5\arcsec$.} 
\label{fig:HoIIstamps1}
\end{figure}

\begin{figure}
\centering
\includegraphics[width=120mm]{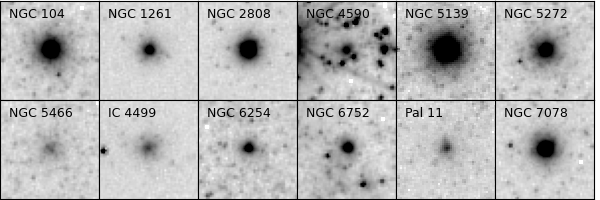}
\caption{Example cut-outs of Milky Way GCs added to the \Euclid\ VIS image of IC\,342. Each cut-out measures $5\arcsec\times5\arcsec$.} 
\label{fig:ic342stamps_mwgc}
\end{figure}

\begin{figure}
\centering
\includegraphics[width=160mm]{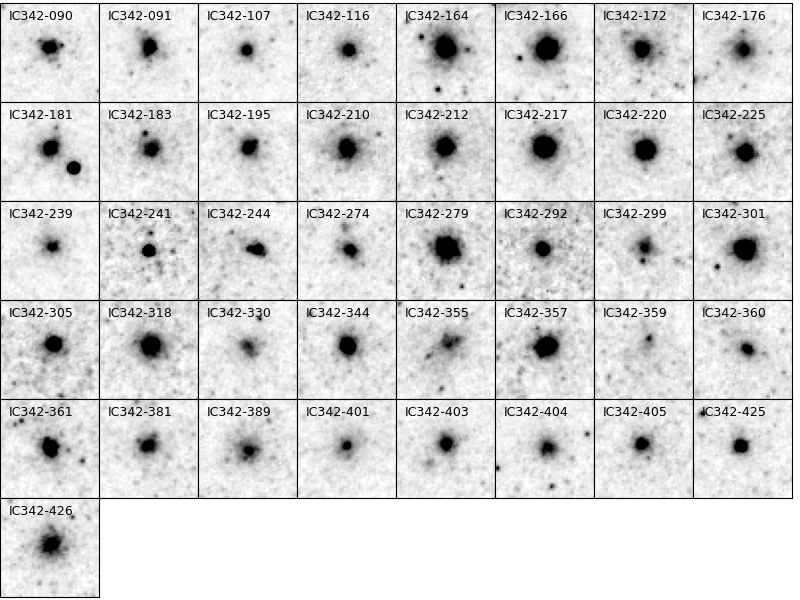}
\caption{HST/ACS F606W cut-outs of the GC candidates in IC\,342. Compared to Fig.~\ref{fig:ic342stamps1gc}, these stamps are zoomed in by a factor of two. Each cut-out thus measures $2\farcs5\times2\farcs5$.} 
\label{fig:hst_ic342stamps1gc}
\label{LastPage}
\end{figure}

\end{appendix}

\end{document}